\documentclass[
 reprint,amsmath,amssymb,aps,
prl,
]{revtex4-2}

\usepackage{graphicx}
\usepackage{dcolumn}
\usepackage{bm}
\usepackage{hyperref}
\hypersetup{colorlinks, linkcolor={blue}, citecolor={blue}, urlcolor={blue}}
\usepackage{xcolor}
\usepackage{pifont}

\begin{document}
\title{Absence of Quantum-Metric-Induced Intrinsic Longitudinal Response}

\author{Ping Tang}
\email{tang.ping.a2@tohoku.ac.jp}
\affiliation{Institute for Materials Research, Tohoku University,
2-1-1 Katahira, Sendai 980-8577, Japan} 
\date{\today}

\begin{abstract}
Nonlinear charge transport in solids has emerged as a powerful probe of the quantum geometric properties of Bloch electrons. While the Berry curvature underlies the intrinsic anomalous Hall effect, recent studies have suggested that the quantum metric may generate both \emph{intrinsic} nonlinear Hall and longitudinal transport. Here, using standard quantum-mechanical perturbation theory, we demonstrate that the quantum-metric-induced intrinsic longitudinal response identically vanishes, even though the corresponding intrinsic Hall response is allowed. This conclusion follows from the dissipationless nature of intrinsic currents and holds independently of band structure details and to all orders in the nonlinear response. Our work resolves existing inconsistencies in the theoretical formulation of quantum-metric-induced nonlinear transport and suggests a reexamination of recently reported intrinsic longitudinal responses attributed to the quantum metric.
\end{abstract}

\maketitle
Nonlinear charge transport in solids has recently attracted tremendous attention as a powerful probe of the quantum geometric properties of electronic wave functions beyond the linear-response regime~\cite{jiang2025revealing, du2021nonlinear}. For instance, the nonlinear Hall conductivity in nonmagnetic systems probes the momentum-resolved Berry curvature through its dipole moment~\cite{PhysRevLett.115.216806,PhysRevLett.121.266601,ma2019observation,kang2019nonlinear,tiwari2021giant,kumar2021room,PhysRevLett.129.186801,he2022graphene,huang2023giant}. In contrast to the well-studied Berry curvature~\cite{berry1984quantal,RevModPhys.82.1959}, the quantum metric ~\cite{provost1980riemannian,PhysRevLett.65.1697}, which measures the distance between Bloch states in momentum space, enters the semiclassical dynamics of electron wave packets at nonlinear order \cite{PhysRevLett.112.166601,PhysRevB.91.214405,PhysRevResearch.4.013217}. In magnetic (time-reversal-breaking) systems, it gives rise to an intrinsic contribution to the nonlinear Hall effect~\cite{PhysRevLett.127.277201,PhysRevLett.127.277202,wang2023quantum,gao2023quantum,yu2025quantum}.

Another important class of nonlinear transport is the longitudinal current response to an applied electric field ($\mathbf{E} \parallel \mathbf{J}$), known as nonreciprocal (longitudinal) transport \cite{ideue2017bulk,tokura2018nonreciprocal,PhysRevLett.122.057206,yasuda2020large,zhao2020magnetic,li2024observation} or magnetochiral anisotropy \cite{PhysRevLett.87.236602,yokouchi2017electrical,PhysRevLett.128.176602}, in which the resistance depends on the current direction. Such responses are typically attributed to second-order Drude conductivities with inversion-asymmetric electron dispersions~\cite{ideue2017bulk,PhysRevResearch.2.043081,PhysRevLett.133.096802}, and promise applications for two-terminal current rectification \cite{ideue2017bulk,tokura2018nonreciprocal} and detection of the N\'eel vector orientation in antiferromagnets~\cite{godinho2018electrically,long2025two,13pd-tlzp}. Beyond nonlinear Hall responses, recent theories~\cite{PhysRevLett.132.026301,PhysRevB.108.L201405} have suggested that the quantum metric can also include an \emph{intrinsic} nonreciprocal conductivity that could dominate over extrinsic Drude contributions in low-mobility systems. Despite this intriguing proposal, a consistent formulation of such intrinsic conductivity remains lacking~\cite{qiang2026clarification}, with different approaches leading to qualitatively conflicting results~\cite{PhysRevLett.112.166601,PhysRevLett.132.026301,PhysRevB.108.L201405}. Remarkably, Wang \textit{et al.}~\cite{wang2023quantum} reported a quantum-metric-induced longitudinal conductivity in MnBi$_2$Te$_4$ that is comparable in magnitude to the induced nonlinear Hall conductivity; however, Gao \textit{et al.}~\cite{gao2023quantum} did not observe such a longitudinal response in the same material. These discrepancies, both theoretical and experimental, call for a systematic and comprehensive investigation of quantum-metric-induced nonlinear transport.

In this Letter, we demonstrate the absence of a quantum-metric-induced intrinsic longitudinal response using the standard quantum-mechanical perturbation theory. The quantum metric is involved in a field-quadratic correction to the electron energy, as well as a redistribution of electrons according to the modified equilibrium state, such that the energy dissipation rate of the induced intrinsic currents vanishes, $\mathbf{E}\cdot\mathbf{J}^{\text{int}}=0$, i.e., the absence of a longitudinal intrinsic response. We distinguish between two types of electron energies: the gauge-\emph{invariant} intrinsic band energy, whose field correction is associated with the polarization work of the electric field, and the gauge-\emph{variant} full energy including the potential energy in the external field. We show that the former is the proper energy entering the semiclassical equations of motion, whereas any improper treatment of the field-induced energy correction or the equilibrium distribution leads to an incorrect finite intrinsic longitudinal response.

Consider the Hamiltonian of electrons in solids subject to a uniform static electric filed, $\hat{H}=\hat{H}_{0}+e\hat{\mathbf{r}}\cdot\mathbf{E}$, where $\hat{H}_{0}$ is the unperturbed crystal Hamiltonian with Bloch eigenstates $\vert \psi_{n}(\mathbf{k}) \rangle = e^{i\mathbf{k}\cdot\hat{\mathbf{r}}} \vert u_{n}(\mathbf{k}) \rangle$, and $\vert u_{n}(\mathbf{k}) \rangle$ denotes the cell-periodic part. The applied electric field modifies both the Bloch wavefunctions and the electron distribution function. The resulting charge current density reads
\begin{equation}
\mathbf{J}=-\frac{e}{V}\sum_{n\mathbf{k}}\tilde{\mathbf{v}}_{n} f_{n}, \label{charge}
\end{equation}
with $e<0$ the electron charge and $V$ the system volume. Here $\tilde{\mathbf{v}}_{n}=\langle\tilde{\psi}_{n}\vert\hat{\mathbf{v}}\vert\tilde{\psi}_{n}\rangle$ is the velocity expectation value evaluated with the field-perturbed eigenstates $\vert\tilde{\psi}_{n}\rangle=e^{i\mathbf{k}\cdot\mathbf{r}}\vert\tilde{u}_{n}\rangle$, with $\hat{\mathbf{v}}=(i\hbar)^{-1}[\hat{\mathbf{r}},\hat{H}]$ the velocity operator; $f_{n}$ is the corresponding distribution function governed by the semiclassical Boltzmann transport equation,
\begin{equation}
\frac{\partial f_{n}}{\partial t}-\frac{e\mathbf{E}}{\hbar}\cdot\frac{\partial f_{n}}{\partial \mathbf{k}}=-\frac{f_{n}-f_{n}^{eq}}{\tau},
\end{equation}
where $\tau$ is the relaxation time and $f_{n}^{eq}$ is the equilibrium distribution of the \emph{perturbed} electrons. In the steady state, 
\begin{equation}
f_{n}=f_{n}^{eq}+\frac{e\tau E_{a}}{\hbar}\frac{\partial f_{n}^{eq}}{\partial k_{a}} +\frac{e^2\tau^2E_{a}E_{b}}{\hbar^2} \frac{\partial^2 f_{n}^{eq}}{\partial k_{a}\partial k_b}+\mathcal{O}(\tau^3), \label{fn}
\end{equation}
where $a,b$ denote Cartesian components, and summation over repeated indices is implied. The electric field not only induces a shift of the distribution in momentum space, but also alters the equilibrium distribution $f_{n}^{eq}$ through field-induced corrections to the electron energy. Here we focus on the second-order nonlinear transport $J_{a}\simeq\sigma_{ab}E_{b}+\sigma_{abc}E_{b}E_{c}$ by expanding the velocity and distribution up to the second order in the electric field, \textit{i.e.,} $\tilde{\mathbf{v}}_{n}\simeq\tilde{\mathbf{v}}_{n}^{(0)}+\tilde{\mathbf{v}}_{n}^{(1)}+\tilde{\mathbf{v}}_{n}^{(2)}$ and $f_{n}\simeq f_{n}^{(0)}+f_{n}^{(1)}+f_{n}^{(2)}$. Using standard second-order perturbation theory, we derive in the Supplemental Material~\cite{SM} the perturbed Bloch eigenstates as
\begin{widetext}
\begin{align}
\vert \tilde{u}_{n}\rangle&=\left\{1-\frac{e^2}{2}E_{a}E_{b}\sum_{m\neq n}\frac{\mathcal{A}_{mn}^{a}\mathcal{A}_{nm}^{b}}{(\varepsilon_{m}-\varepsilon_{n})^2} \right\}\vert u_{n}\rangle +eE_{a}\sum_{m\neq n}\frac{\mathcal{A}_{mn}^{a}}{\varepsilon_{n}-\varepsilon_{m}}\vert u_{m}\rangle\nonumber\\
&+e^2E_{a}E_{b}\sum_{m\neq n}\left\{\sum_{l\neq n}\frac{\mathcal{A}_{ml}^{a}\mathcal{A}_{ln}^{b}}{(\varepsilon_{n}-\varepsilon_{m})(\varepsilon_{n}-\varepsilon_{l})}-\frac{\mathcal{A}_{nn}^{a}\mathcal{A}_{mn}^{b}}{(\varepsilon_{m}-\varepsilon_{n})^2}+\frac{i}{\varepsilon_{n}-\varepsilon_{m}}\left(\frac{\partial}{\partial k_{a}}\frac{\mathcal{A}_{mn}^{b}}{\varepsilon_{n}-\varepsilon_{m}}\right) \right\}\vert u_{m}\rangle+\mathcal{O}(E^3)
\end{align}  
\end{widetext}
where $\varepsilon_{n}$ and $\mathcal{A}_{nm}^{a}=i\langle u_{n}\vert\partial_{k_{a}}\vert u_{m}\rangle$ are the band energy and the generalized Berry connection of the unperturbed Bloch states, respectively. The field-quadratic term in the projection onto $\vert u_{n}\rangle$ follows from the normalization condition $\langle\tilde{u}_{n}\vert\tilde{u}_{n}\rangle=1$. The corresponding velocity expectation value reads~\cite{SM}
\begin{equation}
\tilde{v}_{n}^{a}=\frac{\partial\varepsilon_{n}}{\hbar\partial k_{a}}+\frac{e}{\hbar}E_{b}\Omega_n^{ab} + \frac{e^2}{\hbar} E_b E_c\,\Xi_n^{abc}  \label{velocity}
\end{equation}
where $\Omega_n^{ab}=\partial_{k_a}\mathcal{A}_{n}^{b}-\partial_{k_b}\mathcal{A}_{n}^{a}$ is the unperturbed Berry curvature, and $\Xi_n^{abc}=\partial{k_a}G_{n}^{bc}-(\partial_{k_{b}} G_{n}^{ac} +\partial_{k_{c}} G_{n}^{ab})$ characterizes the filed-quadratic correction, with $G_{n}^{ab}=\sum_{m(\neq n)}\mathrm{Re}[\mathcal{A}_{nm}^{a}\mathcal{A}_{mn}^{b}]/(\varepsilon_{n}-\varepsilon_{m})$ the band-normalized quantum metric~\cite{noteG}. 

\emph{Proper energy correction---}The velocity Eq.~(\ref{velocity}) is derived by standard quantum-mechanical perturbation theory, in which the conventional anomalous velocity induced by the Berry curvature naturally emerges from the fist-order perturbation of the Bloch wavefunctions. To compare with the results from semiclassical wave-packet dynamics~\cite{PhysRevLett.112.166601}, we rewrite Eq.~(\ref{velocity}) in the form 
\begin{equation}
\tilde{v}_{n}^{a}=\frac{\partial\tilde{\varepsilon}_{n}}{\hbar\partial k_{a}}+\frac{e}{\hbar}E_{b}\tilde{\Omega}_{n}^{ab}
\end{equation}
where $\tilde{\varepsilon}_{n}=\varepsilon_{n}-e^2 E_{a}E_{b} G_{n}^{ab}$ and $\tilde{\Omega}_{n}^{ab}=\partial_{k_a}\tilde{\mathcal{A}}_{n}^{b}-\partial_{k_{b}}\tilde{\mathcal{A}}_{n}^{a}$ is the Berry curvature defined in terms of the field-corrected Berry connection $\tilde{\mathcal{A}}_{n}^{a}=i\langle\tilde{u}_{n}\vert \partial_{k_a}\vert\tilde{u}_{n}\rangle=\mathcal{A}_{n}^{a}+2eE_{b}G_{n}^{ab}+\mathcal{O}(E^2)$. Here we identify $\tilde{\varepsilon}_{n}=\langle\tilde{\psi}_{n}\vert \hat{H}_{0} \vert \tilde{\psi}_{n}\rangle$ as the intrinsic, gauge-\emph{invariant} band energy, which differs from the full electron energy under the field
\begin{align}
\tilde{\varepsilon}_{n}^{\prime}=&\langle\tilde{\psi}_{n}\vert\hat{H} \vert\tilde{\psi}_{n}\rangle=\tilde{\varepsilon}_{n}+eE_{a}\tilde{\mathcal{A}}_{n}^{a}\nonumber\\
=&\varepsilon_{n}+eE_{a}\mathcal{A}_{n}^{a}+e^2 E_{a}E_{b} G_{n}^{ab}
\label{energy}
\end{align}
by the external potential energy $eE_{a}\tilde{A}_{n}^{a}$. In contrast to $\tilde{\varepsilon}_{n}$, $\tilde{\varepsilon}_{n}^{\prime}$ includes a gauge-\emph{variant} $E$-linear term that reflects the coordinate-dependent character of the potential energy part, and an $E$-quadratic correction with an \textit{opposite} sign. From Eq.~(\ref{velocity}), $\tilde{\varepsilon}_{n}$, rather than $\tilde{\varepsilon}_{n}^{\prime}$, is the proper field-perturbed electron energy in the equation of motion. In Ref.~\cite{PhysRevLett.132.026301}, however, $\tilde{\varepsilon}_{n}^{\prime}$ was adopted, and its field correction was obtained solely from $eE_{a}\tilde{\mathcal{A}}_{n}^{a}$, while overlooking the correction in $\tilde{\varepsilon}_{n}$, leading to a field-quadratic contribution twice that in Eq.~(\ref{energy}).

The field-induced correction in $\tilde{\varepsilon}_{n}$ is associated with the polarization work of the electric field. Over a differential time interval $dt$, the total work done by the field on electrons reads
\begin{equation}
dW=-e\mathbf{E}\cdot\tilde{\mathbf{v}}_{n}dt=-\frac{e \mathbf{E}}{\hbar}\cdot\frac{\partial\varepsilon_{n}}{\partial \mathbf{k}}dt-e \mathbf{E}\cdot d\tilde{\boldsymbol{\mathcal{A}}}_{n}, \label{W}
\end{equation}
where the first term corresponds to the directional acceleration of electrons and increases their kinetic energy, while the second represents the polarization work due to the induced shift of the wave-packet center (i.e., the perturbation to the electronic wavefunctions)~\cite{notep}. The former is dissipated by disorder scattering into Joule heating, while the latter corrects the intrinsic band energy by
\begin{equation}
\delta \tilde{\varepsilon}_{n}=\int_{0}^{\mathbf{E}}-e\mathbf{E}\cdot d\tilde{\boldsymbol{\mathcal{A}}}_{n}=-e^2E_{a}E_{b} G_{n}^{ab}
\end{equation}
where $d\tilde{\mathcal{A}}_{n}^{a}/dE_{b}=2e G_{n}^{ab}$ corresponds to the Berry-connection polarizability~\cite{PhysRevLett.112.166601,PhysRevB.91.214405}. Consequently, the equilibrium electron distribution follows the Fermi-Dirac function with the field-modified energy $\tilde{\varepsilon}_{n}$~\cite{PhysRevLett.112.166601,qiang2026clarification}, i.e., $f_{n}^{eq}=1/[e^{(\tilde{\varepsilon}_{n}-\mu)/k_{B}T}+1]$, where $\mu$ is the chemical potential. Under the constraint of fixed electron number number, $\mu$ is determined as
\begin{equation}
\mu(\mathbf{E}, T)=\mu_{0}(T)+e^2 \chi_{\mu}^{ab}E_{a}E_{b} 
\end{equation}
where $\mu_{0}$ the unperturbed chemical potential, and the second term represents the field-induced correction, with
\begin{equation}
\chi_{\mu}^{ab}(T)=\frac{\sum_{n\mathbf{k}}G_{n}^{ab}\partial_{\varepsilon_{n}}f_{n}^{(0)}}{\sum_{n\mathbf{k}} \partial_{\varepsilon_{n}}f_{n}^{(0)}}\simeq- G_{\varepsilon_{F}}^{ab},\,\, \text{( $T\rightarrow{0}$)}.
\end{equation}
Here $f_{n}^{(0)}$ is the unperturbed Fermi-Dirac distribution, and $G_{\varepsilon_{F}}^{ab}=D_{F}\sum_{n\mathbf{k}}\delta (\varepsilon_{F}-\varepsilon_{n}) G_{n}^{ab}$ is the energy-resolved quantum metric at the Fermi level $\varepsilon_{F}=\mu_{0}(T\rightarrow0)$, with $D_{F}$ the density of state. Up to second order in the electric field,
\begin{align}
f_{n}^{eq}=&f_{n}^{(0)}-e^2\frac{\partial f_{n}^{(0)}}{\partial \varepsilon_{n}}(G_{n}^{ab}+\chi_{\mu}^{ab})E_{a}E_{b} +\mathcal{O}(E^3), \label{fol}
\end{align}
where the field-induced correction arises from variations in both the electron energy $\delta\tilde{\varepsilon}_{n}$ and the chemical potential $\delta\mu$, such that the total electron number remains unchanged. Eq.~(\ref{fol}) shows that in equilibrium, the electric field also induces a redistribution of electrons according to the corrected energy $\tilde{\varepsilon}_{n}$. This corresponds to a concomitant entropy change of the electron system (at a given temperature $T$) associated with the polarization work performed by the field, and should be distinguished from the nonequilibrium transport arising from the field-induced momentum shift, as described in Eq.~(\ref{fn}). 

\emph{Second-order conductivities---}Substituting Eqs.~(\ref{velocity}) and (\ref{fol}) into Eq.~(\ref{charge}) yields the second-order conductivity as
\begin{align}
\sigma_{abc}=&-\frac{\tau^2e^3}{\hbar^3V}\sum_{n\mathbf{k}}(\partial_{k_{a}}\varepsilon_{n})\partial_{k_{b}}\partial_{k_{c}}f_{n}^{(0)}\nonumber\\
&-\frac{\tau e^3}{2\hbar^2V}\sum_{n\mathbf{k}}\left[\Omega_{n}^{ab}\partial_{k_{c}}f_{n}^{(0)}+\Omega_{n}^{ac}\partial_{k_{b}}f_{n}^{(0)}\right]+\sigma_{abc}^{\text{int}}.
\end{align}
The first ($\sim\tau^2$) and second ($\sim\tau$) terms represent the contribution from second-order Drude conductivity and the Berry curvature dipole, respectively. The remaining contribution, which does not scale with the relaxation time, originates from the intrinsic current carried by the field-modified equilibrium state, $\mathbf{J}^{\text{int}}=-(e/V)\sum_{n\mathbf{k}}\tilde{\mathbf{v}}_{n}f_{n}^{eq}$, yielding
\begin{align}
\sigma_{abc}^{\text{int}}=&-\frac{e^3}{\hbar V}\sum_{n\mathbf{k}}\Big\{\Big[\partial_{k_a}G_{n}^{bc}-\left(\partial_{k_{b}}G_{n}^{ac}+\partial_{k_{c}} G_{n}^{ab}\right)\Big]f_{n}^{(0)} \nonumber\\
&-(G_{n}^{bc}+\chi_{\mu}^{bc})\partial_{k_{a}}f_{n}^{(0)}\Big\} \label{cond}
\end{align}
where the two terms in the curly brackets arise from the field-corrected velocity and the equilibrium distribution, respectively. In the thermodynamic limit $\sum_{\mathbf{k}}\rightarrow V/(2\pi)^d \int d\mathbf{k}$~\cite{Dfa}, and applying integration by parts to the last term, Eq.~(\ref{cond}) reduces to
\begin{align}
\sigma_{abc}^{\text{int}}=&-\frac{e^3}{\hbar}\sum_{n}\int_{\mathbf{k}}\left[2\partial_{k_a}G_{n}^{bc}-\left(\partial_{k_{b}}G_{n}^{ac}+\partial_{k_{c}} G_{n}^{ab}\right)\right]f_{n}^{(0)} \label{con2}
\end{align}
where $\int_{\mathbf{k}}$ denotes the shorthand for $\int d\mathbf{k}/(2\pi)^d$. The contribution from the chemical potential variation vanishes, as it shifts the band-normalized quantum metric only by a $\mathbf{k}$-independent value. Eq.~(\ref{con2}) shows that only transverse (Hall-type) intrinsic currents $(\mathbf{J}^{\text{int}}\perp \mathbf{E})$ are allowed, while the longitudinal component $(\mathbf{J}^{\text{int}}\parallel \mathbf{E})$ is identically absent, since\begin{align}
\mathbf{E}\cdot\mathbf{J}^{\text{int}}&=-\frac{e}{V}\sum_{n\mathbf{k}}\mathbf{E}\cdot\tilde{\mathbf{v}}_{n} f_{n}^{eq}=\sigma_{abc}^{\text{int}}E_{a}E_{b}E_{c}=0. \label{Pd}
\end{align}
Here, the rate of work done by the field $\mathbf{E}\cdot\tilde{\mathbf{v}}_{n}$, as shown in Eq.~(\ref{W}), is compensated by that associated with the redistribution of electrons according to the modified equilibrium state. Eq.~(\ref{Pd}) follows from the \emph{dissipationless} nature of the intrinsic current, i.e., it does not consume power supplied by the field, in contrast to the dissipative (extrinsic) current components arising from the non-equilibrium ($\tau$-dependent) part of the distribution function $\delta f_{n}=f_{n}-f_{n}^{eq}$. 

Finally, we discuss the discrepancies among existing theories. Our result [Eq.~(\ref{con2})] agrees with that derived from semiclassical wave-packet dynamics~\cite{PhysRevLett.112.166601}, but differs qualitatively from those based on the Luttinger-Kohn approach~\cite{PhysRevLett.132.026301} and density-matrix formalism~\cite{PhysRevB.108.L201405}, which predict a finite intrinsic longitudinal conductivity induced by the quantum metric. In Ref.~\cite{PhysRevLett.132.026301}, the field correction to the electron energy is identified from $e\mathbf{E}\cdot\boldsymbol{\tilde{\mathcal{A}}}_{n}$, while neglecting the correction to the equilibrium distribution, thereby yielding a first term twice as large as that in Eq.~(\ref{con2}). In contrast, Ref.~\cite{PhysRevB.108.L201405} assumes a relaxation time of $\tau/2$ when evaluating the field-quadratic correction to the density matrix. This assumption appears insufficiently justified and leads to a second-order Drude conductivity half of the conventional semiclassical Boltzmann result~\cite{PhysRevLett.115.216806,ideue2017bulk,PhysRevResearch.2.043081,PhysRevLett.133.096802}. 

\emph{Conclusion---}We demonstrate that while the quantum metric can generate an intrinsic nonlinear Hall conductivity, the intrinsic longitudinal conductivity identically vanishes. This conclusion is governed by the dissipationless nature of intrinsic currents carried by the field-modified equilibrium state, and therefore holds independently of band structure details and to all orders in the nonlinear response. 
Our findings reveal a fundamental distinction between transverse and longitudinal nonlinear responses, and are consistent with observations of a dominant quantum-metric-induced nonlinear Hall conductivity but a vanishing nonlinear longitudinal response~\cite{PhysRevLett.127.277201,PhysRevLett.127.277202,gao2023quantum,yu2025quantum}. The ``intrinsic" ($\tau$-independent) longitudinal nonlinear signal reported by Wang \textit{et al.}~\cite{wang2023quantum} in low-mobility MnBi$_2$Te$_4$ likely originates from extrinsic disorder-induced mechanisms~\cite{PhysRevB.100.165422,du2019disorder,PhysRevB.100.195117,PhysRevResearch.4.013001,PhysRevB.110.174423}, such as nonlinear side-jump and skew-scattering contributions, where the role of the quantum metric remains largely unexplored~\cite{jiang2025revealing}. For example, it remains an open question whether the quantum metric can induce a $\tau$-independent extrinsic nonlinear conductivity in the presence of disorder, analogous to the extrinsic anomalous Hall effect caused by the side-jump mechanism \cite{RevModPhys.82.1539}.

\emph{Acknowledges---}P. T. acknowledges financial support by JSPS KAKENHI Grant-in-Aid for Scientific Research (B) No. 26K00625.

\bibliography{reference}

\begin{thebibliography}{52}%
\makeatletter
\providecommand \@ifxundefined [1]{%
 \@ifx{#1\undefined}
}%
\providecommand \@ifnum [1]{%
 \ifnum #1\expandafter \@firstoftwo
 \else \expandafter \@secondoftwo
 \fi
}%
\providecommand \@ifx [1]{%
 \ifx #1\expandafter \@firstoftwo
 \else \expandafter \@secondoftwo
 \fi
}%
\providecommand \natexlab [1]{#1}%
\providecommand \enquote  [1]{``#1''}%
\providecommand \bibnamefont  [1]{#1}%
\providecommand \bibfnamefont [1]{#1}%
\providecommand \citenamefont [1]{#1}%
\providecommand \href@noop [0]{\@secondoftwo}%
\providecommand \href [0]{\begingroup \@sanitize@url \@href}%
\providecommand \@href[1]{\@@startlink{#1}\@@href}%
\providecommand \@@href[1]{\endgroup#1\@@endlink}%
\providecommand \@sanitize@url [0]{\catcode `\\12\catcode `\$12\catcode `\&12\catcode `\#12\catcode `\^12\catcode `\_12\catcode `\%12\relax}%
\providecommand \@@startlink[1]{}%
\providecommand \@@endlink[0]{}%
\providecommand \url  [0]{\begingroup\@sanitize@url \@url }%
\providecommand \@url [1]{\endgroup\@href {#1}{\urlprefix }}%
\providecommand \urlprefix  [0]{URL }%
\providecommand \Eprint [0]{\href }%
\providecommand \doibase [0]{https://doi.org/}%
\providecommand \selectlanguage [0]{\@gobble}%
\providecommand \bibinfo  [0]{\@secondoftwo}%
\providecommand \bibfield  [0]{\@secondoftwo}%
\providecommand \translation [1]{[#1]}%
\providecommand \BibitemOpen [0]{}%
\providecommand \bibitemStop [0]{}%
\providecommand \bibitemNoStop [0]{.\EOS\space}%
\providecommand \EOS [0]{\spacefactor3000\relax}%
\providecommand \BibitemShut  [1]{\csname bibitem#1\endcsname}%
\let\auto@bib@innerbib\@empty
\bibitem [{\citenamefont {Jiang}\ \emph {et~al.}(2025)\citenamefont {Jiang}, \citenamefont {Holder},\ and\ \citenamefont {Yan}}]{jiang2025revealing}%
  \BibitemOpen
  \bibfield  {author} {\bibinfo {author} {\bibfnamefont {Y.}~\bibnamefont {Jiang}}, \bibinfo {author} {\bibfnamefont {T.}~\bibnamefont {Holder}},\ and\ \bibinfo {author} {\bibfnamefont {B.}~\bibnamefont {Yan}},\ }\bibfield  {title} {\bibinfo {title} {Revealing quantum geometry in nonlinear quantum materials},\ }\href {https://doi.org/10.1088/1361-6633/ade454} {\bibfield  {journal} {\bibinfo  {journal} {Reports on Progress in Physics}\ }\textbf {\bibinfo {volume} {88}},\ \bibinfo {pages} {076502} (\bibinfo {year} {2025})}\BibitemShut {NoStop}%
\bibitem [{\citenamefont {Du}\ \emph {et~al.}(2021)\citenamefont {Du}, \citenamefont {Lu},\ and\ \citenamefont {Xie}}]{du2021nonlinear}%
  \BibitemOpen
  \bibfield  {author} {\bibinfo {author} {\bibfnamefont {Z.}~\bibnamefont {Du}}, \bibinfo {author} {\bibfnamefont {H.-Z.}\ \bibnamefont {Lu}},\ and\ \bibinfo {author} {\bibfnamefont {X.}~\bibnamefont {Xie}},\ }\bibfield  {title} {\bibinfo {title} {Nonlinear hall effects},\ }\href {https://doi.org/10.1038/s42254-021-00359-6} {\bibfield  {journal} {\bibinfo  {journal} {Nature Reviews Physics}\ }\textbf {\bibinfo {volume} {3}},\ \bibinfo {pages} {744} (\bibinfo {year} {2021})}\BibitemShut {NoStop}%
\bibitem [{\citenamefont {Sodemann}\ and\ \citenamefont {Fu}(2015)}]{PhysRevLett.115.216806}%
  \BibitemOpen
  \bibfield  {author} {\bibinfo {author} {\bibfnamefont {I.}~\bibnamefont {Sodemann}}\ and\ \bibinfo {author} {\bibfnamefont {L.}~\bibnamefont {Fu}},\ }\bibfield  {title} {\bibinfo {title} {Quantum nonlinear hall effect induced by berry curvature dipole in time-reversal invariant materials},\ }\href {https://doi.org/10.1103/PhysRevLett.115.216806} {\bibfield  {journal} {\bibinfo  {journal} {Phys. Rev. Lett.}\ }\textbf {\bibinfo {volume} {115}},\ \bibinfo {pages} {216806} (\bibinfo {year} {2015})}\BibitemShut {NoStop}%
\bibitem [{\citenamefont {Du}\ \emph {et~al.}(2018)\citenamefont {Du}, \citenamefont {Wang}, \citenamefont {Lu},\ and\ \citenamefont {Xie}}]{PhysRevLett.121.266601}%
  \BibitemOpen
  \bibfield  {author} {\bibinfo {author} {\bibfnamefont {Z.~Z.}\ \bibnamefont {Du}}, \bibinfo {author} {\bibfnamefont {C.~M.}\ \bibnamefont {Wang}}, \bibinfo {author} {\bibfnamefont {H.-Z.}\ \bibnamefont {Lu}},\ and\ \bibinfo {author} {\bibfnamefont {X.~C.}\ \bibnamefont {Xie}},\ }\bibfield  {title} {\bibinfo {title} {Band signatures for strong nonlinear hall effect in bilayer ${\mathrm{wte}}_{2}$},\ }\href {https://doi.org/10.1103/PhysRevLett.121.266601} {\bibfield  {journal} {\bibinfo  {journal} {Phys. Rev. Lett.}\ }\textbf {\bibinfo {volume} {121}},\ \bibinfo {pages} {266601} (\bibinfo {year} {2018})}\BibitemShut {NoStop}%
\bibitem [{\citenamefont {Ma}\ \emph {et~al.}(2019)\citenamefont {Ma}, \citenamefont {Xu}, \citenamefont {Shen}, \citenamefont {MacNeill}, \citenamefont {Fatemi}, \citenamefont {Chang}, \citenamefont {Mier~Valdivia}, \citenamefont {Wu}, \citenamefont {Du}, \citenamefont {Hsu} \emph {et~al.}}]{ma2019observation}%
  \BibitemOpen
  \bibfield  {author} {\bibinfo {author} {\bibfnamefont {Q.}~\bibnamefont {Ma}}, \bibinfo {author} {\bibfnamefont {S.-Y.}\ \bibnamefont {Xu}}, \bibinfo {author} {\bibfnamefont {H.}~\bibnamefont {Shen}}, \bibinfo {author} {\bibfnamefont {D.}~\bibnamefont {MacNeill}}, \bibinfo {author} {\bibfnamefont {V.}~\bibnamefont {Fatemi}}, \bibinfo {author} {\bibfnamefont {T.-R.}\ \bibnamefont {Chang}}, \bibinfo {author} {\bibfnamefont {A.~M.}\ \bibnamefont {Mier~Valdivia}}, \bibinfo {author} {\bibfnamefont {S.}~\bibnamefont {Wu}}, \bibinfo {author} {\bibfnamefont {Z.}~\bibnamefont {Du}}, \bibinfo {author} {\bibfnamefont {C.-H.}\ \bibnamefont {Hsu}}, \emph {et~al.},\ }\bibfield  {title} {\bibinfo {title} {Observation of the nonlinear hall effect under time-reversal-symmetric conditions},\ }\href {https://doi.org/10.1038/s41586-018-0807-6} {\bibfield  {journal} {\bibinfo  {journal} {Nature}\ }\textbf {\bibinfo {volume} {565}},\ \bibinfo {pages} {337} (\bibinfo {year} {2019})}\BibitemShut {NoStop}%
\bibitem [{\citenamefont {Kang}\ \emph {et~al.}(2019)\citenamefont {Kang}, \citenamefont {Li}, \citenamefont {Sohn}, \citenamefont {Shan},\ and\ \citenamefont {Mak}}]{kang2019nonlinear}%
  \BibitemOpen
  \bibfield  {author} {\bibinfo {author} {\bibfnamefont {K.}~\bibnamefont {Kang}}, \bibinfo {author} {\bibfnamefont {T.}~\bibnamefont {Li}}, \bibinfo {author} {\bibfnamefont {E.}~\bibnamefont {Sohn}}, \bibinfo {author} {\bibfnamefont {J.}~\bibnamefont {Shan}},\ and\ \bibinfo {author} {\bibfnamefont {K.~F.}\ \bibnamefont {Mak}},\ }\bibfield  {title} {\bibinfo {title} {Nonlinear anomalous hall effect in few-layer wte2},\ }\href {https://doi.org/10.1038/s41563-019-0294-7} {\bibfield  {journal} {\bibinfo  {journal} {Nature Materials}\ }\textbf {\bibinfo {volume} {18}},\ \bibinfo {pages} {324} (\bibinfo {year} {2019})}\BibitemShut {NoStop}%
\bibitem [{\citenamefont {Tiwari}\ \emph {et~al.}(2021)\citenamefont {Tiwari}, \citenamefont {Chen}, \citenamefont {Zhong}, \citenamefont {Drueke}, \citenamefont {Koo}, \citenamefont {Kaczmarek}, \citenamefont {Xiao}, \citenamefont {Gao}, \citenamefont {Luo}, \citenamefont {Niu} \emph {et~al.}}]{tiwari2021giant}%
  \BibitemOpen
  \bibfield  {author} {\bibinfo {author} {\bibfnamefont {A.}~\bibnamefont {Tiwari}}, \bibinfo {author} {\bibfnamefont {F.}~\bibnamefont {Chen}}, \bibinfo {author} {\bibfnamefont {S.}~\bibnamefont {Zhong}}, \bibinfo {author} {\bibfnamefont {E.}~\bibnamefont {Drueke}}, \bibinfo {author} {\bibfnamefont {J.}~\bibnamefont {Koo}}, \bibinfo {author} {\bibfnamefont {A.}~\bibnamefont {Kaczmarek}}, \bibinfo {author} {\bibfnamefont {C.}~\bibnamefont {Xiao}}, \bibinfo {author} {\bibfnamefont {J.}~\bibnamefont {Gao}}, \bibinfo {author} {\bibfnamefont {X.}~\bibnamefont {Luo}}, \bibinfo {author} {\bibfnamefont {Q.}~\bibnamefont {Niu}}, \emph {et~al.},\ }\bibfield  {title} {\bibinfo {title} {Giant c-axis nonlinear anomalous hall effect in td-mote2 and wte2},\ }\href {https://doi.org/10.1038/s41467-021-22343-5} {\bibfield  {journal} {\bibinfo  {journal} {Nature communications}\ }\textbf {\bibinfo {volume} {12}},\ \bibinfo {pages} {2049} (\bibinfo {year} {2021})}\BibitemShut {NoStop}%
\bibitem [{\citenamefont {Kumar}\ \emph {et~al.}(2021)\citenamefont {Kumar}, \citenamefont {Hsu}, \citenamefont {Sharma}, \citenamefont {Chang}, \citenamefont {Yu}, \citenamefont {Wang}, \citenamefont {Eda}, \citenamefont {Liang},\ and\ \citenamefont {Yang}}]{kumar2021room}%
  \BibitemOpen
  \bibfield  {author} {\bibinfo {author} {\bibfnamefont {D.}~\bibnamefont {Kumar}}, \bibinfo {author} {\bibfnamefont {C.-H.}\ \bibnamefont {Hsu}}, \bibinfo {author} {\bibfnamefont {R.}~\bibnamefont {Sharma}}, \bibinfo {author} {\bibfnamefont {T.-R.}\ \bibnamefont {Chang}}, \bibinfo {author} {\bibfnamefont {P.}~\bibnamefont {Yu}}, \bibinfo {author} {\bibfnamefont {J.}~\bibnamefont {Wang}}, \bibinfo {author} {\bibfnamefont {G.}~\bibnamefont {Eda}}, \bibinfo {author} {\bibfnamefont {G.}~\bibnamefont {Liang}},\ and\ \bibinfo {author} {\bibfnamefont {H.}~\bibnamefont {Yang}},\ }\bibfield  {title} {\bibinfo {title} {Room-temperature nonlinear hall effect and wireless radiofrequency rectification in weyl semimetal tairte4},\ }\href {https://doi.org/10.1038/s41565-020-00839-3} {\bibfield  {journal} {\bibinfo  {journal} {Nature Nanotechnology}\ }\textbf {\bibinfo {volume} {16}},\ \bibinfo {pages} {421} (\bibinfo {year} {2021})}\BibitemShut {NoStop}%
\bibitem [{\citenamefont {Duan}\ \emph {et~al.}(2022)\citenamefont {Duan}, \citenamefont {Jian}, \citenamefont {Gao}, \citenamefont {Peng}, \citenamefont {Zhong}, \citenamefont {Feng}, \citenamefont {Mao},\ and\ \citenamefont {Yao}}]{PhysRevLett.129.186801}%
  \BibitemOpen
  \bibfield  {author} {\bibinfo {author} {\bibfnamefont {J.}~\bibnamefont {Duan}}, \bibinfo {author} {\bibfnamefont {Y.}~\bibnamefont {Jian}}, \bibinfo {author} {\bibfnamefont {Y.}~\bibnamefont {Gao}}, \bibinfo {author} {\bibfnamefont {H.}~\bibnamefont {Peng}}, \bibinfo {author} {\bibfnamefont {J.}~\bibnamefont {Zhong}}, \bibinfo {author} {\bibfnamefont {Q.}~\bibnamefont {Feng}}, \bibinfo {author} {\bibfnamefont {J.}~\bibnamefont {Mao}},\ and\ \bibinfo {author} {\bibfnamefont {Y.}~\bibnamefont {Yao}},\ }\bibfield  {title} {\bibinfo {title} {Giant second-order nonlinear hall effect in twisted bilayer graphene},\ }\href {https://doi.org/10.1103/PhysRevLett.129.186801} {\bibfield  {journal} {\bibinfo  {journal} {Phys. Rev. Lett.}\ }\textbf {\bibinfo {volume} {129}},\ \bibinfo {pages} {186801} (\bibinfo {year} {2022})}\BibitemShut {NoStop}%
\bibitem [{\citenamefont {He}\ \emph {et~al.}(2022)\citenamefont {He}, \citenamefont {Koon}, \citenamefont {Isobe}, \citenamefont {Tan}, \citenamefont {Hu}, \citenamefont {Neto}, \citenamefont {Fu},\ and\ \citenamefont {Yang}}]{he2022graphene}%
  \BibitemOpen
  \bibfield  {author} {\bibinfo {author} {\bibfnamefont {P.}~\bibnamefont {He}}, \bibinfo {author} {\bibfnamefont {G.~K.~W.}\ \bibnamefont {Koon}}, \bibinfo {author} {\bibfnamefont {H.}~\bibnamefont {Isobe}}, \bibinfo {author} {\bibfnamefont {J.~Y.}\ \bibnamefont {Tan}}, \bibinfo {author} {\bibfnamefont {J.}~\bibnamefont {Hu}}, \bibinfo {author} {\bibfnamefont {A.~H.~C.}\ \bibnamefont {Neto}}, \bibinfo {author} {\bibfnamefont {L.}~\bibnamefont {Fu}},\ and\ \bibinfo {author} {\bibfnamefont {H.}~\bibnamefont {Yang}},\ }\bibfield  {title} {\bibinfo {title} {Graphene moir{\'e} superlattices with giant quantum nonlinearity of chiral bloch electrons},\ }\href {https://doi.org/10.1038/s41565-021-01060-6} {\bibfield  {journal} {\bibinfo  {journal} {Nature Nanotechnology}\ }\textbf {\bibinfo {volume} {17}},\ \bibinfo {pages} {378} (\bibinfo {year} {2022})}\BibitemShut {NoStop}%
\bibitem [{\citenamefont {Huang}\ \emph {et~al.}(2023)\citenamefont {Huang}, \citenamefont {Wu}, \citenamefont {Hu}, \citenamefont {Cai}, \citenamefont {Li}, \citenamefont {An}, \citenamefont {Feng}, \citenamefont {Ye}, \citenamefont {Lin}, \citenamefont {Law} \emph {et~al.}}]{huang2023giant}%
  \BibitemOpen
  \bibfield  {author} {\bibinfo {author} {\bibfnamefont {M.}~\bibnamefont {Huang}}, \bibinfo {author} {\bibfnamefont {Z.}~\bibnamefont {Wu}}, \bibinfo {author} {\bibfnamefont {J.}~\bibnamefont {Hu}}, \bibinfo {author} {\bibfnamefont {X.}~\bibnamefont {Cai}}, \bibinfo {author} {\bibfnamefont {E.}~\bibnamefont {Li}}, \bibinfo {author} {\bibfnamefont {L.}~\bibnamefont {An}}, \bibinfo {author} {\bibfnamefont {X.}~\bibnamefont {Feng}}, \bibinfo {author} {\bibfnamefont {Z.}~\bibnamefont {Ye}}, \bibinfo {author} {\bibfnamefont {N.}~\bibnamefont {Lin}}, \bibinfo {author} {\bibfnamefont {K.~T.}\ \bibnamefont {Law}}, \emph {et~al.},\ }\bibfield  {title} {\bibinfo {title} {Giant nonlinear hall effect in twisted bilayer wse2},\ }\href {https://doi.org/10.1093/nsr/nwac232} {\bibfield  {journal} {\bibinfo  {journal} {National Science Review}\ }\textbf {\bibinfo {volume} {10}},\ \bibinfo {pages} {nwac232} (\bibinfo {year} {2023})}\BibitemShut {NoStop}%
\bibitem [{\citenamefont {Berry}(1984)}]{berry1984quantal}%
  \BibitemOpen
  \bibfield  {author} {\bibinfo {author} {\bibfnamefont {M.~V.}\ \bibnamefont {Berry}},\ }\bibfield  {title} {\bibinfo {title} {Quantal phase factors accompanying adiabatic changes},\ }\href {https://doi.org/10.1098/rspa.1984.0023} {\bibfield  {journal} {\bibinfo  {journal} {Proceedings of the Royal Society of London. A. Mathematical and Physical Sciences}\ }\textbf {\bibinfo {volume} {392}},\ \bibinfo {pages} {45} (\bibinfo {year} {1984})}\BibitemShut {NoStop}%
\bibitem [{\citenamefont {Xiao}\ \emph {et~al.}(2010)\citenamefont {Xiao}, \citenamefont {Chang},\ and\ \citenamefont {Niu}}]{RevModPhys.82.1959}%
  \BibitemOpen
  \bibfield  {author} {\bibinfo {author} {\bibfnamefont {D.}~\bibnamefont {Xiao}}, \bibinfo {author} {\bibfnamefont {M.-C.}\ \bibnamefont {Chang}},\ and\ \bibinfo {author} {\bibfnamefont {Q.}~\bibnamefont {Niu}},\ }\bibfield  {title} {\bibinfo {title} {Berry phase effects on electronic properties},\ }\href {https://doi.org/10.1103/RevModPhys.82.1959} {\bibfield  {journal} {\bibinfo  {journal} {Rev. Mod. Phys.}\ }\textbf {\bibinfo {volume} {82}},\ \bibinfo {pages} {1959} (\bibinfo {year} {2010})}\BibitemShut {NoStop}%
\bibitem [{\citenamefont {Provost}\ and\ \citenamefont {Vallee}(1980)}]{provost1980riemannian}%
  \BibitemOpen
  \bibfield  {author} {\bibinfo {author} {\bibfnamefont {J.}~\bibnamefont {Provost}}\ and\ \bibinfo {author} {\bibfnamefont {G.}~\bibnamefont {Vallee}},\ }\bibfield  {title} {\bibinfo {title} {Riemannian structure on manifolds of quantum states},\ }\href {https://doi.org/10.1007/BF02193559} {\bibfield  {journal} {\bibinfo  {journal} {Communications in Mathematical Physics}\ }\textbf {\bibinfo {volume} {76}},\ \bibinfo {pages} {289} (\bibinfo {year} {1980})}\BibitemShut {NoStop}%
\bibitem [{\citenamefont {Anandan}\ and\ \citenamefont {Aharonov}(1990)}]{PhysRevLett.65.1697}%
  \BibitemOpen
  \bibfield  {author} {\bibinfo {author} {\bibfnamefont {J.}~\bibnamefont {Anandan}}\ and\ \bibinfo {author} {\bibfnamefont {Y.}~\bibnamefont {Aharonov}},\ }\bibfield  {title} {\bibinfo {title} {Geometry of quantum evolution},\ }\href {https://doi.org/10.1103/PhysRevLett.65.1697} {\bibfield  {journal} {\bibinfo  {journal} {Phys. Rev. Lett.}\ }\textbf {\bibinfo {volume} {65}},\ \bibinfo {pages} {1697} (\bibinfo {year} {1990})}\BibitemShut {NoStop}%
\bibitem [{\citenamefont {Gao}\ \emph {et~al.}(2014)\citenamefont {Gao}, \citenamefont {Yang},\ and\ \citenamefont {Niu}}]{PhysRevLett.112.166601}%
  \BibitemOpen
  \bibfield  {author} {\bibinfo {author} {\bibfnamefont {Y.}~\bibnamefont {Gao}}, \bibinfo {author} {\bibfnamefont {S.~A.}\ \bibnamefont {Yang}},\ and\ \bibinfo {author} {\bibfnamefont {Q.}~\bibnamefont {Niu}},\ }\bibfield  {title} {\bibinfo {title} {Field induced positional shift of bloch electrons and its dynamical implications},\ }\href {https://doi.org/10.1103/PhysRevLett.112.166601} {\bibfield  {journal} {\bibinfo  {journal} {Phys. Rev. Lett.}\ }\textbf {\bibinfo {volume} {112}},\ \bibinfo {pages} {166601} (\bibinfo {year} {2014})}\BibitemShut {NoStop}%
\bibitem [{\citenamefont {Gao}\ \emph {et~al.}(2015)\citenamefont {Gao}, \citenamefont {Yang},\ and\ \citenamefont {Niu}}]{PhysRevB.91.214405}%
  \BibitemOpen
  \bibfield  {author} {\bibinfo {author} {\bibfnamefont {Y.}~\bibnamefont {Gao}}, \bibinfo {author} {\bibfnamefont {S.~A.}\ \bibnamefont {Yang}},\ and\ \bibinfo {author} {\bibfnamefont {Q.}~\bibnamefont {Niu}},\ }\bibfield  {title} {\bibinfo {title} {Geometrical effects in orbital magnetic susceptibility},\ }\href {https://doi.org/10.1103/PhysRevB.91.214405} {\bibfield  {journal} {\bibinfo  {journal} {Phys. Rev. B}\ }\textbf {\bibinfo {volume} {91}},\ \bibinfo {pages} {214405} (\bibinfo {year} {2015})}\BibitemShut {NoStop}%
\bibitem [{\citenamefont {Smith}\ \emph {et~al.}(2022)\citenamefont {Smith}, \citenamefont {Pullasseri},\ and\ \citenamefont {Srivastava}}]{PhysRevResearch.4.013217}%
  \BibitemOpen
  \bibfield  {author} {\bibinfo {author} {\bibfnamefont {T.~B.}\ \bibnamefont {Smith}}, \bibinfo {author} {\bibfnamefont {L.}~\bibnamefont {Pullasseri}},\ and\ \bibinfo {author} {\bibfnamefont {A.}~\bibnamefont {Srivastava}},\ }\bibfield  {title} {\bibinfo {title} {Momentum-space gravity from the quantum geometry and entropy of bloch electrons},\ }\href {https://doi.org/10.1103/PhysRevResearch.4.013217} {\bibfield  {journal} {\bibinfo  {journal} {Phys. Rev. Res.}\ }\textbf {\bibinfo {volume} {4}},\ \bibinfo {pages} {013217} (\bibinfo {year} {2022})}\BibitemShut {NoStop}%
\bibitem [{\citenamefont {Wang}\ \emph {et~al.}(2021)\citenamefont {Wang}, \citenamefont {Gao},\ and\ \citenamefont {Xiao}}]{PhysRevLett.127.277201}%
  \BibitemOpen
  \bibfield  {author} {\bibinfo {author} {\bibfnamefont {C.}~\bibnamefont {Wang}}, \bibinfo {author} {\bibfnamefont {Y.}~\bibnamefont {Gao}},\ and\ \bibinfo {author} {\bibfnamefont {D.}~\bibnamefont {Xiao}},\ }\bibfield  {title} {\bibinfo {title} {Intrinsic nonlinear hall effect in antiferromagnetic tetragonal cumnas},\ }\href {https://doi.org/10.1103/PhysRevLett.127.277201} {\bibfield  {journal} {\bibinfo  {journal} {Phys. Rev. Lett.}\ }\textbf {\bibinfo {volume} {127}},\ \bibinfo {pages} {277201} (\bibinfo {year} {2021})}\BibitemShut {NoStop}%
\bibitem [{\citenamefont {Liu}\ \emph {et~al.}(2021)\citenamefont {Liu}, \citenamefont {Zhao}, \citenamefont {Huang}, \citenamefont {Wu}, \citenamefont {Sheng}, \citenamefont {Xiao},\ and\ \citenamefont {Yang}}]{PhysRevLett.127.277202}%
  \BibitemOpen
  \bibfield  {author} {\bibinfo {author} {\bibfnamefont {H.}~\bibnamefont {Liu}}, \bibinfo {author} {\bibfnamefont {J.}~\bibnamefont {Zhao}}, \bibinfo {author} {\bibfnamefont {Y.-X.}\ \bibnamefont {Huang}}, \bibinfo {author} {\bibfnamefont {W.}~\bibnamefont {Wu}}, \bibinfo {author} {\bibfnamefont {X.-L.}\ \bibnamefont {Sheng}}, \bibinfo {author} {\bibfnamefont {C.}~\bibnamefont {Xiao}},\ and\ \bibinfo {author} {\bibfnamefont {S.~A.}\ \bibnamefont {Yang}},\ }\bibfield  {title} {\bibinfo {title} {Intrinsic second-order anomalous hall effect and its application in compensated antiferromagnets},\ }\href {https://doi.org/10.1103/PhysRevLett.127.277202} {\bibfield  {journal} {\bibinfo  {journal} {Phys. Rev. Lett.}\ }\textbf {\bibinfo {volume} {127}},\ \bibinfo {pages} {277202} (\bibinfo {year} {2021})}\BibitemShut {NoStop}%
\bibitem [{\citenamefont {Wang}\ \emph {et~al.}(2023)\citenamefont {Wang}, \citenamefont {Kaplan}, \citenamefont {Zhang}, \citenamefont {Holder}, \citenamefont {Cao}, \citenamefont {Wang}, \citenamefont {Zhou}, \citenamefont {Zhou}, \citenamefont {Jiang}, \citenamefont {Zhang} \emph {et~al.}}]{wang2023quantum}%
  \BibitemOpen
  \bibfield  {author} {\bibinfo {author} {\bibfnamefont {N.}~\bibnamefont {Wang}}, \bibinfo {author} {\bibfnamefont {D.}~\bibnamefont {Kaplan}}, \bibinfo {author} {\bibfnamefont {Z.}~\bibnamefont {Zhang}}, \bibinfo {author} {\bibfnamefont {T.}~\bibnamefont {Holder}}, \bibinfo {author} {\bibfnamefont {N.}~\bibnamefont {Cao}}, \bibinfo {author} {\bibfnamefont {A.}~\bibnamefont {Wang}}, \bibinfo {author} {\bibfnamefont {X.}~\bibnamefont {Zhou}}, \bibinfo {author} {\bibfnamefont {F.}~\bibnamefont {Zhou}}, \bibinfo {author} {\bibfnamefont {Z.}~\bibnamefont {Jiang}}, \bibinfo {author} {\bibfnamefont {C.}~\bibnamefont {Zhang}}, \emph {et~al.},\ }\bibfield  {title} {\bibinfo {title} {Quantum-metric-induced nonlinear transport in a topological antiferromagnet},\ }\href {https://doi.org/10.1038/s41586-023-06363-3} {\bibfield  {journal} {\bibinfo  {journal} {Nature}\ }\textbf {\bibinfo {volume} {621}},\ \bibinfo {pages} {487} (\bibinfo {year} {2023})}\BibitemShut {NoStop}%
\bibitem [{\citenamefont {Gao}\ \emph {et~al.}(2023)\citenamefont {Gao}, \citenamefont {Liu}, \citenamefont {Qiu}, \citenamefont {Ghosh}, \citenamefont {V.~Trevisan}, \citenamefont {Onishi}, \citenamefont {Hu}, \citenamefont {Qian}, \citenamefont {Tien}, \citenamefont {Chen} \emph {et~al.}}]{gao2023quantum}%
  \BibitemOpen
  \bibfield  {author} {\bibinfo {author} {\bibfnamefont {A.}~\bibnamefont {Gao}}, \bibinfo {author} {\bibfnamefont {Y.-F.}\ \bibnamefont {Liu}}, \bibinfo {author} {\bibfnamefont {J.-X.}\ \bibnamefont {Qiu}}, \bibinfo {author} {\bibfnamefont {B.}~\bibnamefont {Ghosh}}, \bibinfo {author} {\bibfnamefont {T.}~\bibnamefont {V.~Trevisan}}, \bibinfo {author} {\bibfnamefont {Y.}~\bibnamefont {Onishi}}, \bibinfo {author} {\bibfnamefont {C.}~\bibnamefont {Hu}}, \bibinfo {author} {\bibfnamefont {T.}~\bibnamefont {Qian}}, \bibinfo {author} {\bibfnamefont {H.-J.}\ \bibnamefont {Tien}}, \bibinfo {author} {\bibfnamefont {S.-W.}\ \bibnamefont {Chen}}, \emph {et~al.},\ }\bibfield  {title} {\bibinfo {title} {Quantum metric nonlinear hall effect in a topological antiferromagnetic heterostructure},\ }\href {https://doi.org/10.1126/science.adf1506} {\bibfield  {journal} {\bibinfo  {journal} {Science}\ }\textbf {\bibinfo {volume} {381}},\ \bibinfo {pages} {181} (\bibinfo {year} {2023})}\BibitemShut {NoStop}%
\bibitem [{\citenamefont {Yu}\ \emph {et~al.}(2025)\citenamefont {Yu}, \citenamefont {Li}, \citenamefont {Bie}, \citenamefont {Yan}, \citenamefont {Zhou}, \citenamefont {Yu},\ and\ \citenamefont {Yang}}]{yu2025quantum}%
  \BibitemOpen
  \bibfield  {author} {\bibinfo {author} {\bibfnamefont {H.}~\bibnamefont {Yu}}, \bibinfo {author} {\bibfnamefont {X.}~\bibnamefont {Li}}, \bibinfo {author} {\bibfnamefont {Y.-Q.}\ \bibnamefont {Bie}}, \bibinfo {author} {\bibfnamefont {L.}~\bibnamefont {Yan}}, \bibinfo {author} {\bibfnamefont {L.}~\bibnamefont {Zhou}}, \bibinfo {author} {\bibfnamefont {P.}~\bibnamefont {Yu}},\ and\ \bibinfo {author} {\bibfnamefont {G.}~\bibnamefont {Yang}},\ }\bibfield  {title} {\bibinfo {title} {Quantum metric third-order nonlinear hall effect in a non-centrosymmetric ferromagnet},\ }\href {https://doi.org/10.1038/s41467-025-63133-7} {\bibfield  {journal} {\bibinfo  {journal} {Nature Communications}\ }\textbf {\bibinfo {volume} {16}},\ \bibinfo {pages} {7698} (\bibinfo {year} {2025})}\BibitemShut {NoStop}%
\bibitem [{\citenamefont {Ideue}\ \emph {et~al.}(2017)\citenamefont {Ideue}, \citenamefont {Hamamoto}, \citenamefont {Koshikawa}, \citenamefont {Ezawa}, \citenamefont {Shimizu}, \citenamefont {Kaneko}, \citenamefont {Tokura}, \citenamefont {Nagaosa},\ and\ \citenamefont {Iwasa}}]{ideue2017bulk}%
  \BibitemOpen
  \bibfield  {author} {\bibinfo {author} {\bibfnamefont {T.}~\bibnamefont {Ideue}}, \bibinfo {author} {\bibfnamefont {K.}~\bibnamefont {Hamamoto}}, \bibinfo {author} {\bibfnamefont {S.}~\bibnamefont {Koshikawa}}, \bibinfo {author} {\bibfnamefont {M.}~\bibnamefont {Ezawa}}, \bibinfo {author} {\bibfnamefont {S.}~\bibnamefont {Shimizu}}, \bibinfo {author} {\bibfnamefont {Y.}~\bibnamefont {Kaneko}}, \bibinfo {author} {\bibfnamefont {Y.}~\bibnamefont {Tokura}}, \bibinfo {author} {\bibfnamefont {N.}~\bibnamefont {Nagaosa}},\ and\ \bibinfo {author} {\bibfnamefont {Y.}~\bibnamefont {Iwasa}},\ }\bibfield  {title} {\bibinfo {title} {Bulk rectification effect in a polar semiconductor},\ }\href {https://doi.org/10.1038/nphys4056} {\bibfield  {journal} {\bibinfo  {journal} {Nature Physics}\ }\textbf {\bibinfo {volume} {13}},\ \bibinfo {pages} {578} (\bibinfo {year} {2017})}\BibitemShut {NoStop}%
\bibitem [{\citenamefont {Tokura}\ and\ \citenamefont {Nagaosa}(2018)}]{tokura2018nonreciprocal}%
  \BibitemOpen
  \bibfield  {author} {\bibinfo {author} {\bibfnamefont {Y.}~\bibnamefont {Tokura}}\ and\ \bibinfo {author} {\bibfnamefont {N.}~\bibnamefont {Nagaosa}},\ }\bibfield  {title} {\bibinfo {title} {Nonreciprocal responses from non-centrosymmetric quantum materials},\ }\href {https://doi.org/10.1038/s41467-018-05759-4} {\bibfield  {journal} {\bibinfo  {journal} {Nature Communications}\ }\textbf {\bibinfo {volume} {9}},\ \bibinfo {pages} {3740} (\bibinfo {year} {2018})}\BibitemShut {NoStop}%
\bibitem [{\citenamefont {Aoki}\ \emph {et~al.}(2019)\citenamefont {Aoki}, \citenamefont {Kousaka},\ and\ \citenamefont {Togawa}}]{PhysRevLett.122.057206}%
  \BibitemOpen
  \bibfield  {author} {\bibinfo {author} {\bibfnamefont {R.}~\bibnamefont {Aoki}}, \bibinfo {author} {\bibfnamefont {Y.}~\bibnamefont {Kousaka}},\ and\ \bibinfo {author} {\bibfnamefont {Y.}~\bibnamefont {Togawa}},\ }\bibfield  {title} {\bibinfo {title} {Anomalous nonreciprocal electrical transport on chiral magnetic order},\ }\href {https://doi.org/10.1103/PhysRevLett.122.057206} {\bibfield  {journal} {\bibinfo  {journal} {Phys. Rev. Lett.}\ }\textbf {\bibinfo {volume} {122}},\ \bibinfo {pages} {057206} (\bibinfo {year} {2019})}\BibitemShut {NoStop}%
\bibitem [{\citenamefont {Yasuda}\ \emph {et~al.}(2020)\citenamefont {Yasuda}, \citenamefont {Morimoto}, \citenamefont {Yoshimi}, \citenamefont {Mogi}, \citenamefont {Tsukazaki}, \citenamefont {Kawamura}, \citenamefont {Takahashi}, \citenamefont {Kawasaki}, \citenamefont {Nagaosa},\ and\ \citenamefont {Tokura}}]{yasuda2020large}%
  \BibitemOpen
  \bibfield  {author} {\bibinfo {author} {\bibfnamefont {K.}~\bibnamefont {Yasuda}}, \bibinfo {author} {\bibfnamefont {T.}~\bibnamefont {Morimoto}}, \bibinfo {author} {\bibfnamefont {R.}~\bibnamefont {Yoshimi}}, \bibinfo {author} {\bibfnamefont {M.}~\bibnamefont {Mogi}}, \bibinfo {author} {\bibfnamefont {A.}~\bibnamefont {Tsukazaki}}, \bibinfo {author} {\bibfnamefont {M.}~\bibnamefont {Kawamura}}, \bibinfo {author} {\bibfnamefont {K.~S.}\ \bibnamefont {Takahashi}}, \bibinfo {author} {\bibfnamefont {M.}~\bibnamefont {Kawasaki}}, \bibinfo {author} {\bibfnamefont {N.}~\bibnamefont {Nagaosa}},\ and\ \bibinfo {author} {\bibfnamefont {Y.}~\bibnamefont {Tokura}},\ }\bibfield  {title} {\bibinfo {title} {Large non-reciprocal charge transport mediated by quantum anomalous hall edge states},\ }\href {https://doi.org/10.1038/s41565-020-0733-2} {\bibfield  {journal} {\bibinfo  {journal} {Nature Nanotechnology}\ }\textbf {\bibinfo {volume} {15}},\ \bibinfo {pages} {831} (\bibinfo {year} {2020})}\BibitemShut {NoStop}%
\bibitem [{\citenamefont {Zhao}\ \emph {et~al.}(2020)\citenamefont {Zhao}, \citenamefont {Fei}, \citenamefont {Song}, \citenamefont {Choi}, \citenamefont {Palomaki}, \citenamefont {Sun}, \citenamefont {Malinowski}, \citenamefont {McGuire}, \citenamefont {Chu}, \citenamefont {Xu} \emph {et~al.}}]{zhao2020magnetic}%
  \BibitemOpen
  \bibfield  {author} {\bibinfo {author} {\bibfnamefont {W.}~\bibnamefont {Zhao}}, \bibinfo {author} {\bibfnamefont {Z.}~\bibnamefont {Fei}}, \bibinfo {author} {\bibfnamefont {T.}~\bibnamefont {Song}}, \bibinfo {author} {\bibfnamefont {H.~K.}\ \bibnamefont {Choi}}, \bibinfo {author} {\bibfnamefont {T.}~\bibnamefont {Palomaki}}, \bibinfo {author} {\bibfnamefont {B.}~\bibnamefont {Sun}}, \bibinfo {author} {\bibfnamefont {P.}~\bibnamefont {Malinowski}}, \bibinfo {author} {\bibfnamefont {M.~A.}\ \bibnamefont {McGuire}}, \bibinfo {author} {\bibfnamefont {J.-H.}\ \bibnamefont {Chu}}, \bibinfo {author} {\bibfnamefont {X.}~\bibnamefont {Xu}}, \emph {et~al.},\ }\bibfield  {title} {\bibinfo {title} {Magnetic proximity and nonreciprocal current switching in a monolayer wte2 helical edge},\ }\href {https://doi.org/10.1038/s41563-020-0620-0} {\bibfield  {journal} {\bibinfo  {journal} {Nature Materials}\ }\textbf {\bibinfo {volume} {19}},\ \bibinfo {pages} {503} (\bibinfo {year} {2020})}\BibitemShut {NoStop}%
\bibitem [{\citenamefont {Li}\ \emph {et~al.}(2024)\citenamefont {Li}, \citenamefont {Wang}, \citenamefont {Zhang}, \citenamefont {Qin}, \citenamefont {Ying}, \citenamefont {Wei}, \citenamefont {Dai}, \citenamefont {Guo}, \citenamefont {Chen}, \citenamefont {Zhang} \emph {et~al.}}]{li2024observation}%
  \BibitemOpen
  \bibfield  {author} {\bibinfo {author} {\bibfnamefont {C.}~\bibnamefont {Li}}, \bibinfo {author} {\bibfnamefont {R.}~\bibnamefont {Wang}}, \bibinfo {author} {\bibfnamefont {S.}~\bibnamefont {Zhang}}, \bibinfo {author} {\bibfnamefont {Y.}~\bibnamefont {Qin}}, \bibinfo {author} {\bibfnamefont {Z.}~\bibnamefont {Ying}}, \bibinfo {author} {\bibfnamefont {B.}~\bibnamefont {Wei}}, \bibinfo {author} {\bibfnamefont {Z.}~\bibnamefont {Dai}}, \bibinfo {author} {\bibfnamefont {F.}~\bibnamefont {Guo}}, \bibinfo {author} {\bibfnamefont {W.}~\bibnamefont {Chen}}, \bibinfo {author} {\bibfnamefont {R.}~\bibnamefont {Zhang}}, \emph {et~al.},\ }\bibfield  {title} {\bibinfo {title} {Observation of giant non-reciprocal charge transport from quantum hall states in a topological insulator},\ }\href {https://doi.org/10.1038/s41563-024-01874-4} {\bibfield  {journal} {\bibinfo  {journal} {Nature Materials}\ }\textbf {\bibinfo {volume} {23}},\ \bibinfo {pages} {1208} (\bibinfo {year} {2024})}\BibitemShut {NoStop}%
\bibitem [{\citenamefont {Rikken}\ \emph {et~al.}(2001)\citenamefont {Rikken}, \citenamefont {F\"olling},\ and\ \citenamefont {Wyder}}]{PhysRevLett.87.236602}%
  \BibitemOpen
  \bibfield  {author} {\bibinfo {author} {\bibfnamefont {G.~L. J.~A.}\ \bibnamefont {Rikken}}, \bibinfo {author} {\bibfnamefont {J.}~\bibnamefont {F\"olling}},\ and\ \bibinfo {author} {\bibfnamefont {P.}~\bibnamefont {Wyder}},\ }\bibfield  {title} {\bibinfo {title} {Electrical magnetochiral anisotropy},\ }\href {https://doi.org/10.1103/PhysRevLett.87.236602} {\bibfield  {journal} {\bibinfo  {journal} {Phys. Rev. Lett.}\ }\textbf {\bibinfo {volume} {87}},\ \bibinfo {pages} {236602} (\bibinfo {year} {2001})}\BibitemShut {NoStop}%
\bibitem [{\citenamefont {Yokouchi}\ \emph {et~al.}(2017)\citenamefont {Yokouchi}, \citenamefont {Kanazawa}, \citenamefont {Kikkawa}, \citenamefont {Morikawa}, \citenamefont {Shibata}, \citenamefont {Arima}, \citenamefont {Taguchi}, \citenamefont {Kagawa},\ and\ \citenamefont {Tokura}}]{yokouchi2017electrical}%
  \BibitemOpen
  \bibfield  {author} {\bibinfo {author} {\bibfnamefont {T.}~\bibnamefont {Yokouchi}}, \bibinfo {author} {\bibfnamefont {N.}~\bibnamefont {Kanazawa}}, \bibinfo {author} {\bibfnamefont {A.}~\bibnamefont {Kikkawa}}, \bibinfo {author} {\bibfnamefont {D.}~\bibnamefont {Morikawa}}, \bibinfo {author} {\bibfnamefont {K.}~\bibnamefont {Shibata}}, \bibinfo {author} {\bibfnamefont {T.}~\bibnamefont {Arima}}, \bibinfo {author} {\bibfnamefont {Y.}~\bibnamefont {Taguchi}}, \bibinfo {author} {\bibfnamefont {F.}~\bibnamefont {Kagawa}},\ and\ \bibinfo {author} {\bibfnamefont {Y.}~\bibnamefont {Tokura}},\ }\bibfield  {title} {\bibinfo {title} {Electrical magnetochiral effect induced by chiral spin fluctuations},\ }\href {https://doi.org/10.1038/s41467-017-01094-2} {\bibfield  {journal} {\bibinfo  {journal} {Nature Communications}\ }\textbf {\bibinfo {volume} {8}},\ \bibinfo {pages} {866} (\bibinfo {year} {2017})}\BibitemShut {NoStop}%
\bibitem [{\citenamefont {Wang}\ \emph {et~al.}(2022)\citenamefont {Wang}, \citenamefont {Legg}, \citenamefont {B\"omerich}, \citenamefont {Park}, \citenamefont {Biesenkamp}, \citenamefont {Taskin}, \citenamefont {Braden}, \citenamefont {Rosch},\ and\ \citenamefont {Ando}}]{PhysRevLett.128.176602}%
  \BibitemOpen
  \bibfield  {author} {\bibinfo {author} {\bibfnamefont {Y.}~\bibnamefont {Wang}}, \bibinfo {author} {\bibfnamefont {H.~F.}\ \bibnamefont {Legg}}, \bibinfo {author} {\bibfnamefont {T.}~\bibnamefont {B\"omerich}}, \bibinfo {author} {\bibfnamefont {J.}~\bibnamefont {Park}}, \bibinfo {author} {\bibfnamefont {S.}~\bibnamefont {Biesenkamp}}, \bibinfo {author} {\bibfnamefont {A.~A.}\ \bibnamefont {Taskin}}, \bibinfo {author} {\bibfnamefont {M.}~\bibnamefont {Braden}}, \bibinfo {author} {\bibfnamefont {A.}~\bibnamefont {Rosch}},\ and\ \bibinfo {author} {\bibfnamefont {Y.}~\bibnamefont {Ando}},\ }\bibfield  {title} {\bibinfo {title} {Gigantic magnetochiral anisotropy in the topological semimetal ${\mathrm{zrte}}_{5}$},\ }\href {https://doi.org/10.1103/PhysRevLett.128.176602} {\bibfield  {journal} {\bibinfo  {journal} {Phys. Rev. Lett.}\ }\textbf {\bibinfo {volume} {128}},\ \bibinfo {pages} {176602} (\bibinfo {year} {2022})}\BibitemShut {NoStop}%
\bibitem [{\citenamefont {Watanabe}\ and\ \citenamefont {Yanase}(2020)}]{PhysRevResearch.2.043081}%
  \BibitemOpen
  \bibfield  {author} {\bibinfo {author} {\bibfnamefont {H.}~\bibnamefont {Watanabe}}\ and\ \bibinfo {author} {\bibfnamefont {Y.}~\bibnamefont {Yanase}},\ }\bibfield  {title} {\bibinfo {title} {Nonlinear electric transport in odd-parity magnetic multipole systems: Application to mn-based compounds},\ }\href {https://doi.org/10.1103/PhysRevResearch.2.043081} {\bibfield  {journal} {\bibinfo  {journal} {Phys. Rev. Res.}\ }\textbf {\bibinfo {volume} {2}},\ \bibinfo {pages} {043081} (\bibinfo {year} {2020})}\BibitemShut {NoStop}%
\bibitem [{\citenamefont {Zhao}\ \emph {et~al.}(2024)\citenamefont {Zhao}, \citenamefont {Tao}, \citenamefont {Fu}, \citenamefont {Bellaiche},\ and\ \citenamefont {Ma}}]{PhysRevLett.133.096802}%
  \BibitemOpen
  \bibfield  {author} {\bibinfo {author} {\bibfnamefont {H.~J.}\ \bibnamefont {Zhao}}, \bibinfo {author} {\bibfnamefont {L.}~\bibnamefont {Tao}}, \bibinfo {author} {\bibfnamefont {Y.}~\bibnamefont {Fu}}, \bibinfo {author} {\bibfnamefont {L.}~\bibnamefont {Bellaiche}},\ and\ \bibinfo {author} {\bibfnamefont {Y.}~\bibnamefont {Ma}},\ }\bibfield  {title} {\bibinfo {title} {General theory for longitudinal nonreciprocal charge transport},\ }\href {https://doi.org/10.1103/PhysRevLett.133.096802} {\bibfield  {journal} {\bibinfo  {journal} {Phys. Rev. Lett.}\ }\textbf {\bibinfo {volume} {133}},\ \bibinfo {pages} {096802} (\bibinfo {year} {2024})}\BibitemShut {NoStop}%
\bibitem [{\citenamefont {Godinho}\ \emph {et~al.}(2018)\citenamefont {Godinho}, \citenamefont {Reichlov{\'a}}, \citenamefont {Kriegner}, \citenamefont {Nov{\'a}k}, \citenamefont {Olejn{\'\i}k}, \citenamefont {Ka{\v{s}}par}, \citenamefont {{\v{S}}ob{\'a}{\v{n}}}, \citenamefont {Wadley}, \citenamefont {Campion}, \citenamefont {Otxoa} \emph {et~al.}}]{godinho2018electrically}%
  \BibitemOpen
  \bibfield  {author} {\bibinfo {author} {\bibfnamefont {J.}~\bibnamefont {Godinho}}, \bibinfo {author} {\bibfnamefont {H.}~\bibnamefont {Reichlov{\'a}}}, \bibinfo {author} {\bibfnamefont {D.}~\bibnamefont {Kriegner}}, \bibinfo {author} {\bibfnamefont {V.}~\bibnamefont {Nov{\'a}k}}, \bibinfo {author} {\bibfnamefont {K.}~\bibnamefont {Olejn{\'\i}k}}, \bibinfo {author} {\bibfnamefont {Z.}~\bibnamefont {Ka{\v{s}}par}}, \bibinfo {author} {\bibfnamefont {Z.}~\bibnamefont {{\v{S}}ob{\'a}{\v{n}}}}, \bibinfo {author} {\bibfnamefont {P.}~\bibnamefont {Wadley}}, \bibinfo {author} {\bibfnamefont {R.}~\bibnamefont {Campion}}, \bibinfo {author} {\bibfnamefont {R.}~\bibnamefont {Otxoa}}, \emph {et~al.},\ }\bibfield  {title} {\bibinfo {title} {Electrically induced and detected n{\'e}el vector reversal in a collinear antiferromagnet},\ }\href {https://doi.org/10.1038/s41467-018-07092-2} {\bibfield  {journal} {\bibinfo  {journal} {Nature Communications}\ }\textbf {\bibinfo {volume} {9}},\ \bibinfo {pages} {4686} (\bibinfo
  {year} {2018})}\BibitemShut {NoStop}%
\bibitem [{\citenamefont {Long}\ \emph {et~al.}(2025)\citenamefont {Long}, \citenamefont {Zeng}, \citenamefont {Pan}, \citenamefont {Duan},\ and\ \citenamefont {Huang}}]{long2025two}%
  \BibitemOpen
  \bibfield  {author} {\bibinfo {author} {\bibfnamefont {G.}~\bibnamefont {Long}}, \bibinfo {author} {\bibfnamefont {H.}~\bibnamefont {Zeng}}, \bibinfo {author} {\bibfnamefont {M.}~\bibnamefont {Pan}}, \bibinfo {author} {\bibfnamefont {W.}~\bibnamefont {Duan}},\ and\ \bibinfo {author} {\bibfnamefont {H.}~\bibnamefont {Huang}},\ }\bibfield  {title} {\bibinfo {title} {Two-terminal electrical detection of the n{\'e}el vector via longitudinal antiferromagnetic nonreciprocal transport},\ }\href {https://doi.org/10.1021/acs.nanolett.5c02968} {\bibfield  {journal} {\bibinfo  {journal} {Nano Letters}\ }\textbf {\bibinfo {volume} {25}},\ \bibinfo {pages} {14817} (\bibinfo {year} {2025})}\BibitemShut {NoStop}%
\bibitem [{\citenamefont {Sudo}\ \emph {et~al.}(2026)\citenamefont {Sudo}, \citenamefont {Yanagi}, \citenamefont {Akaki}, \citenamefont {Tanida},\ and\ \citenamefont {Kimata}}]{13pd-tlzp}%
  \BibitemOpen
  \bibfield  {author} {\bibinfo {author} {\bibfnamefont {K.}~\bibnamefont {Sudo}}, \bibinfo {author} {\bibfnamefont {Y.}~\bibnamefont {Yanagi}}, \bibinfo {author} {\bibfnamefont {M.}~\bibnamefont {Akaki}}, \bibinfo {author} {\bibfnamefont {H.}~\bibnamefont {Tanida}},\ and\ \bibinfo {author} {\bibfnamefont {M.}~\bibnamefont {Kimata}},\ }\bibfield  {title} {\bibinfo {title} {Large spontaneous nonreciprocal charge transport in a zero-magnetization antiferromagnet},\ }\href {https://doi.org/10.1103/13pd-tlzp} {\bibfield  {journal} {\bibinfo  {journal} {Phys. Rev. Lett.}\ }\textbf {\bibinfo {volume} {136}},\ \bibinfo {pages} {016503} (\bibinfo {year} {2026})}\BibitemShut {NoStop}%
\bibitem [{\citenamefont {Kaplan}\ \emph {et~al.}(2024)\citenamefont {Kaplan}, \citenamefont {Holder},\ and\ \citenamefont {Yan}}]{PhysRevLett.132.026301}%
  \BibitemOpen
  \bibfield  {author} {\bibinfo {author} {\bibfnamefont {D.}~\bibnamefont {Kaplan}}, \bibinfo {author} {\bibfnamefont {T.}~\bibnamefont {Holder}},\ and\ \bibinfo {author} {\bibfnamefont {B.}~\bibnamefont {Yan}},\ }\bibfield  {title} {\bibinfo {title} {Unification of nonlinear anomalous hall effect and nonreciprocal magnetoresistance in metals by the quantum geometry},\ }\href {https://doi.org/10.1103/PhysRevLett.132.026301} {\bibfield  {journal} {\bibinfo  {journal} {Phys. Rev. Lett.}\ }\textbf {\bibinfo {volume} {132}},\ \bibinfo {pages} {026301} (\bibinfo {year} {2024})}\BibitemShut {NoStop}%
\bibitem [{\citenamefont {Das}\ \emph {et~al.}(2023)\citenamefont {Das}, \citenamefont {Lahiri}, \citenamefont {Atencia}, \citenamefont {Culcer},\ and\ \citenamefont {Agarwal}}]{PhysRevB.108.L201405}%
  \BibitemOpen
  \bibfield  {author} {\bibinfo {author} {\bibfnamefont {K.}~\bibnamefont {Das}}, \bibinfo {author} {\bibfnamefont {S.}~\bibnamefont {Lahiri}}, \bibinfo {author} {\bibfnamefont {R.~B.}\ \bibnamefont {Atencia}}, \bibinfo {author} {\bibfnamefont {D.}~\bibnamefont {Culcer}},\ and\ \bibinfo {author} {\bibfnamefont {A.}~\bibnamefont {Agarwal}},\ }\bibfield  {title} {\bibinfo {title} {Intrinsic nonlinear conductivities induced by the quantum metric},\ }\href {https://doi.org/10.1103/PhysRevB.108.L201405} {\bibfield  {journal} {\bibinfo  {journal} {Phys. Rev. B}\ }\textbf {\bibinfo {volume} {108}},\ \bibinfo {pages} {L201405} (\bibinfo {year} {2023})}\BibitemShut {NoStop}%
\bibitem [{\citenamefont {Qiang}\ \emph {et~al.}(2026)\citenamefont {Qiang}, \citenamefont {Liu}, \citenamefont {Gao}, \citenamefont {Lu},\ and\ \citenamefont {Xie}}]{qiang2026clarification}%
  \BibitemOpen
  \bibfield  {author} {\bibinfo {author} {\bibfnamefont {X.-B.}\ \bibnamefont {Qiang}}, \bibinfo {author} {\bibfnamefont {T.}~\bibnamefont {Liu}}, \bibinfo {author} {\bibfnamefont {Z.-X.}\ \bibnamefont {Gao}}, \bibinfo {author} {\bibfnamefont {H.-Z.}\ \bibnamefont {Lu}},\ and\ \bibinfo {author} {\bibfnamefont {X.}~\bibnamefont {Xie}},\ }\bibfield  {title} {\bibinfo {title} {A clarification on quantum-metric-induced nonlinear transport},\ }\href {https://doi.org/10.1002/advs.202514818} {\bibfield  {journal} {\bibinfo  {journal} {Advanced Science}\ }\textbf {\bibinfo {volume} {13}},\ \bibinfo {pages} {e14818} (\bibinfo {year} {2026})}\BibitemShut {NoStop}%
\bibitem [{SM()}]{SM}%
  \BibitemOpen
  \href@noop {} {}\bibinfo {note} {{See the Supplemental Material for the detailed derivation of the perturbed Bloch eigensates and the associated velocity expectation value}}\BibitemShut {NoStop}%
\bibitem [{\citenamefont {{The band-normalized quantum metric is defined as the real part of the band-normalized quantum geometric tensor $Q_{n}^{ab}=\sum_{m\neq n} \mathcal{A}_{nm}^{a}\mathcal{A}_{mn}^{b}/(\varepsilon_{n}-\varepsilon_{m})$}}()}]{noteG}%
  \BibitemOpen
  \bibfield  {author} {\bibinfo {author} {\bibnamefont {{The band-normalized quantum metric is defined as the real part of the band-normalized quantum geometric tensor $Q_{n}^{ab}=\sum_{m\neq n} \mathcal{A}_{nm}^{a}\mathcal{A}_{mn}^{b}/(\varepsilon_{n}-\varepsilon_{m})$}}},\ }\href@noop {} {}\bibinfo {note} {{which is smaller by a factor of two compared to the definition in Ref.~\cite{PhysRevLett.132.026301}}}\BibitemShut {NoStop}%
\bibitem [{not()}]{notep}%
  \BibitemOpen
  \href@noop {} {}\bibinfo {note} {{The polarization work done by an external field reads $dW_{p}=\mathbf{E}\cdot d\mathbf{P}$, where $\mathbf{P}=-e\tilde{\boldsymbol{\mathcal{A}}_{n}}$ represents the electric polarization in the modern theory \cite{RevModPhys.66.899} for an electron in the $n$th perturbed eigenstate}}\BibitemShut {NoStop}%
\bibitem [{\citenamefont {{In the thermodynamic limit}}()}]{Dfa}%
  \BibitemOpen
  \bibfield  {author} {\bibinfo {author} {\bibnamefont {{In the thermodynamic limit}}},\ }\href@noop {} {}\bibinfo {note} {{$\sum_{\mathbf{k}}\rightarrow V/(2\pi)^{d} \int d\mathbf{k} [1+e\mathbf{B}\cdot\boldsymbol{\Omega}_{n}/\hbar]$~\cite{PhysRevLett.95.137204}, where the correction to the phase-space density of states vanishes in the absence of an applied magnetic field $\mathbf{B}$}}\BibitemShut {NoStop}%
\bibitem [{\citenamefont {Xiao}\ \emph {et~al.}(2019)\citenamefont {Xiao}, \citenamefont {Du},\ and\ \citenamefont {Niu}}]{PhysRevB.100.165422}%
  \BibitemOpen
  \bibfield  {author} {\bibinfo {author} {\bibfnamefont {C.}~\bibnamefont {Xiao}}, \bibinfo {author} {\bibfnamefont {Z.~Z.}\ \bibnamefont {Du}},\ and\ \bibinfo {author} {\bibfnamefont {Q.}~\bibnamefont {Niu}},\ }\bibfield  {title} {\bibinfo {title} {Theory of nonlinear hall effects: Modified semiclassics from quantum kinetics},\ }\href {https://doi.org/10.1103/PhysRevB.100.165422} {\bibfield  {journal} {\bibinfo  {journal} {Phys. Rev. B}\ }\textbf {\bibinfo {volume} {100}},\ \bibinfo {pages} {165422} (\bibinfo {year} {2019})}\BibitemShut {NoStop}%
\bibitem [{\citenamefont {Du}\ \emph {et~al.}(2019)\citenamefont {Du}, \citenamefont {Wang}, \citenamefont {Li}, \citenamefont {Lu},\ and\ \citenamefont {Xie}}]{du2019disorder}%
  \BibitemOpen
  \bibfield  {author} {\bibinfo {author} {\bibfnamefont {Z.}~\bibnamefont {Du}}, \bibinfo {author} {\bibfnamefont {C.}~\bibnamefont {Wang}}, \bibinfo {author} {\bibfnamefont {S.}~\bibnamefont {Li}}, \bibinfo {author} {\bibfnamefont {H.-Z.}\ \bibnamefont {Lu}},\ and\ \bibinfo {author} {\bibfnamefont {X.}~\bibnamefont {Xie}},\ }\bibfield  {title} {\bibinfo {title} {Disorder-induced nonlinear hall effect with time-reversal symmetry},\ }\href {https://doi.org/10.1038/s41467-019-10941-3} {\bibfield  {journal} {\bibinfo  {journal} {Nature Communications}\ }\textbf {\bibinfo {volume} {10}},\ \bibinfo {pages} {3047} (\bibinfo {year} {2019})}\BibitemShut {NoStop}%
\bibitem [{\citenamefont {Nandy}\ and\ \citenamefont {Sodemann}(2019)}]{PhysRevB.100.195117}%
  \BibitemOpen
  \bibfield  {author} {\bibinfo {author} {\bibfnamefont {S.}~\bibnamefont {Nandy}}\ and\ \bibinfo {author} {\bibfnamefont {I.}~\bibnamefont {Sodemann}},\ }\bibfield  {title} {\bibinfo {title} {Symmetry and quantum kinetics of the nonlinear hall effect},\ }\href {https://doi.org/10.1103/PhysRevB.100.195117} {\bibfield  {journal} {\bibinfo  {journal} {Phys. Rev. B}\ }\textbf {\bibinfo {volume} {100}},\ \bibinfo {pages} {195117} (\bibinfo {year} {2019})}\BibitemShut {NoStop}%
\bibitem [{\citenamefont {Atencia}\ \emph {et~al.}(2022)\citenamefont {Atencia}, \citenamefont {Niu},\ and\ \citenamefont {Culcer}}]{PhysRevResearch.4.013001}%
  \BibitemOpen
  \bibfield  {author} {\bibinfo {author} {\bibfnamefont {R.~B.}\ \bibnamefont {Atencia}}, \bibinfo {author} {\bibfnamefont {Q.}~\bibnamefont {Niu}},\ and\ \bibinfo {author} {\bibfnamefont {D.}~\bibnamefont {Culcer}},\ }\bibfield  {title} {\bibinfo {title} {Semiclassical response of disordered conductors: Extrinsic carrier velocity and spin and field-corrected collision integral},\ }\href {https://doi.org/10.1103/PhysRevResearch.4.013001} {\bibfield  {journal} {\bibinfo  {journal} {Phys. Rev. Res.}\ }\textbf {\bibinfo {volume} {4}},\ \bibinfo {pages} {013001} (\bibinfo {year} {2022})}\BibitemShut {NoStop}%
\bibitem [{\citenamefont {Mehraeen}(2024)}]{PhysRevB.110.174423}%
  \BibitemOpen
  \bibfield  {author} {\bibinfo {author} {\bibfnamefont {M.}~\bibnamefont {Mehraeen}},\ }\bibfield  {title} {\bibinfo {title} {Quantum kinetic theory of quadratic responses},\ }\href {https://doi.org/10.1103/PhysRevB.110.174423} {\bibfield  {journal} {\bibinfo  {journal} {Phys. Rev. B}\ }\textbf {\bibinfo {volume} {110}},\ \bibinfo {pages} {174423} (\bibinfo {year} {2024})}\BibitemShut {NoStop}%
\bibitem [{\citenamefont {Nagaosa}\ \emph {et~al.}(2010)\citenamefont {Nagaosa}, \citenamefont {Sinova}, \citenamefont {Onoda}, \citenamefont {MacDonald},\ and\ \citenamefont {Ong}}]{RevModPhys.82.1539}%
  \BibitemOpen
  \bibfield  {author} {\bibinfo {author} {\bibfnamefont {N.}~\bibnamefont {Nagaosa}}, \bibinfo {author} {\bibfnamefont {J.}~\bibnamefont {Sinova}}, \bibinfo {author} {\bibfnamefont {S.}~\bibnamefont {Onoda}}, \bibinfo {author} {\bibfnamefont {A.~H.}\ \bibnamefont {MacDonald}},\ and\ \bibinfo {author} {\bibfnamefont {N.~P.}\ \bibnamefont {Ong}},\ }\bibfield  {title} {\bibinfo {title} {Anomalous hall effect},\ }\href {https://doi.org/10.1103/RevModPhys.82.1539} {\bibfield  {journal} {\bibinfo  {journal} {Rev. Mod. Phys.}\ }\textbf {\bibinfo {volume} {82}},\ \bibinfo {pages} {1539} (\bibinfo {year} {2010})}\BibitemShut {NoStop}%
\bibitem [{\citenamefont {Resta}(1994)}]{RevModPhys.66.899}%
  \BibitemOpen
  \bibfield  {author} {\bibinfo {author} {\bibfnamefont {R.}~\bibnamefont {Resta}},\ }\bibfield  {title} {\bibinfo {title} {Macroscopic polarization in crystalline dielectrics: the geometric phase approach},\ }\href {https://doi.org/10.1103/RevModPhys.66.899} {\bibfield  {journal} {\bibinfo  {journal} {Rev. Mod. Phys.}\ }\textbf {\bibinfo {volume} {66}},\ \bibinfo {pages} {899} (\bibinfo {year} {1994})}\BibitemShut {NoStop}%
\bibitem [{\citenamefont {Xiao}\ \emph {et~al.}(2005)\citenamefont {Xiao}, \citenamefont {Shi},\ and\ \citenamefont {Niu}}]{PhysRevLett.95.137204}%
  \BibitemOpen
  \bibfield  {author} {\bibinfo {author} {\bibfnamefont {D.}~\bibnamefont {Xiao}}, \bibinfo {author} {\bibfnamefont {J.}~\bibnamefont {Shi}},\ and\ \bibinfo {author} {\bibfnamefont {Q.}~\bibnamefont {Niu}},\ }\bibfield  {title} {\bibinfo {title} {Berry phase correction to electron density of states in solids},\ }\href {https://doi.org/10.1103/PhysRevLett.95.137204} {\bibfield  {journal} {\bibinfo  {journal} {Phys. Rev. Lett.}\ }\textbf {\bibinfo {volume} {95}},\ \bibinfo {pages} {137204} (\bibinfo {year} {2005})}\BibitemShut {NoStop}%
\end{thebibliography}%
\end{document}


\title{Supplemental Material for ``Absence of Quantum-Metric-Induced Intrinsic Longitudinal Response"}

\author{Ping Tang}
\email{tang.ping.a2@tohoku.ac.jp}	
\affiliation{Institute for Materials Research, Tohoku University, 2-1-1 Katahira, Sendai 980-8577, Japan}
\date{\today}

\maketitle

In the Supplemental Material, we derive the field-perturbed Bloch states, energies, and velocity expectation values to second order in the external field.

\section{Field-perturbed Bloch states and energies}
Consider the Hamiltonian $\hat{H}=\hat{H}_0+e\hat{\mathbf r}\cdot\mathbf E$, where $\hat{H}_0$ is the crystal Hamiltonian satisfying
$\hat{H}_0|\psi_n(\mathbf k)\rangle=\varepsilon_n(\mathbf{k})|\psi_n(\mathbf k)\rangle$, with $\varepsilon_n(\mathbf{k})$ and $|\psi_n(\mathbf k)\rangle$ denoting the unperturbed electron energies and Bloch eigenstates, respectively. Using standard second-order perturbation theory, the normalized field-perturbed eigenstate reads
\begin{align}
|\tilde{\psi}_\mu\rangle
&=
\left[
1-\frac{1}{2}
\sum_{\mu^{\prime}\neq \mu}
\frac{|\hat{V}_{\mu^{\prime}\mu}|^2}
{(\varepsilon_\mu-\varepsilon_{\mu^{\prime}})^2}
\right]|\psi_{\mu}\rangle
+
\sum_{\mu^{\prime}\neq \mu}
\frac{\hat{V}_{\mu^{\prime}\mu}}{\varepsilon_\mu-\varepsilon_{\mu^{\prime}}}|\psi_{\mu^{\prime}}\rangle
\nonumber\\
&\quad
+
\sum_{\mu^{\prime}\neq \mu}
\left[
\sum_{\mu^{\prime\prime}\neq \mu}
\frac{\hat{V}_{\mu^{\prime}\mu^{\prime\prime}}\hat{V}_{\mu^{\prime\prime}\mu}}
{(\varepsilon_\mu-\varepsilon_{\mu^{\prime}})(\varepsilon_\mu-\varepsilon_{\mu^{\prime\prime}})}
-
\frac{\hat{V}_{\mu\mu}\hat{V}_{\mu^{\prime}\mu}}
{(\varepsilon_\mu-\varepsilon_{\mu^{\prime}})^2}
\right]
|\psi_{\mu^{\prime}}\rangle
+\mathcal O(V^3). \label{psi}
\end{align}
where $\hat{V}=e\hat{\mathbf r}\cdot\mathbf E$ is the external perturbation and $\mu=(n,\mathbf k)$ labels the eigenstates. The matrix elements of the perturbation are given by $\hat{V}_{\mu^{\prime}\mu}
=\langle\psi_{n'}(\mathbf k')|e\hat{\mathbf r}\cdot\mathbf E|\psi_n(\mathbf k)\rangle
=e\mathbf E\cdot\mathbf r_{n'n}^{\mathbf k'\mathbf k}$, where
\begin{align}
\mathbf r_{n'n}^{\mathbf k'\mathbf k}
=\langle\psi_{n'}(\mathbf k')|\hat{\mathbf r}|\psi_n(\mathbf k)\rangle=i\delta_{nn'}\frac{\partial}{\partial\mathbf k'}\delta_{\mathbf k\mathbf k'}
+\delta_{\mathbf k\mathbf k'}\boldsymbol{\mathcal A}_{n'n}(\mathbf k)
\label{rs}
\end{align}
is the matrix element of the position operator and $\boldsymbol{\mathcal A}_{n'n}(\mathbf k)
=i\langle u_{n'}(\mathbf k)|\partial_{\mathbf k}|u_n(\mathbf k)\rangle$ denotes the generalized (unperturbed) Berry connection. Substituting $\hat{V}_{\mu^{\prime}\mu}
=e\mathbf E\cdot\mathbf r_{n'n}^{\mathbf k'\mathbf k}$ into Eq.~(\ref{psi}) yields
\begin{align}
\vert \tilde{\psi}_{n}(\mathbf{k})\rangle&=\left\{1-\frac{e^2}{2}\sum_{m\neq n}\frac{(\mathbf{E}\cdot\boldsymbol{\mathcal{A}}_{mn})(\mathbf{E}\cdot\boldsymbol{\mathcal{A}}_{nm})}{(\varepsilon_{m}-\varepsilon_{n})^2} \right\}\vert \psi_{n}(\mathbf{k})\rangle +\sum_{m\neq n}\frac{e\mathbf{E}\cdot\boldsymbol{\mathcal{A}}_{mn}}{\varepsilon_{n}-\varepsilon_{m}}\vert \psi_{m}(\mathbf{k})\rangle\nonumber\\
&+e^2\sum_{m\neq n}\left\{\sum_{l\neq n}\frac{(\mathbf{E}\cdot\boldsymbol{\mathcal{A}}_{ml})(\mathbf{E}\cdot\boldsymbol{\mathcal{A}}_{ln})}{(\varepsilon_{n}-\varepsilon_{m})(\varepsilon_{n}-\varepsilon_{l})}-\frac{(\mathbf{E}\cdot\boldsymbol{\mathcal{A}}_{nn})(\mathbf{E}\cdot\boldsymbol{\mathcal{A}}_{mn})}{(\varepsilon_{m}-\varepsilon_{n})^2} \right\}\vert \psi_{m}(\mathbf{k})\rangle\nonumber\\
&+ie^2\sum_{m\neq n}\frac{\mathbf{E}}{\varepsilon_{n}-\varepsilon_{m}}\cdot\left(\frac{\partial}{\partial \mathbf{k}}\frac{\mathbf{E}\cdot\boldsymbol{\mathcal{A}}_{mn}}{\varepsilon_{n}-\varepsilon_{m}}\right)\vert \psi_{m}(\mathbf{k})\rangle+\mathcal{O}(E^3) \label{psi2}
\end{align} 
which gives the perturbed cell-periodic Bloch function in the main text as
$|\tilde{u}_{n}(\mathbf{k})\rangle=e^{-i\mathbf{k}\cdot\mathbf{r}}|\tilde{\psi}_{n}(\mathbf{k})\rangle$. The last term in Eq.~(\ref{psi2}) originates from the first derivative term in the position matrix element in Eq.~(\ref{rs}). The corresponding electron energy is given by
\begin{equation}
\langle\tilde{\psi}_{n}(\mathbf{k})\vert\hat{H}\vert\tilde{\psi}_{n}(\mathbf{k})\rangle =\langle\tilde{\psi}_{n}(\mathbf{k})\vert\hat{H}_{0}\vert\tilde{\psi}_{n}(\mathbf{k})\rangle +e\mathbf{E}\cdot\boldsymbol{\mathcal{A}}_{n}(\mathbf{k}) \label{en}
\end{equation}
where $\langle\tilde{\psi}_{n}(\mathbf{k})\vert\hat{H}_{0}\vert\tilde{\psi}_{n}(\mathbf{k})\rangle=\varepsilon_{n}-e^2E_{a}E_{b}G_{n}^{ab}$ is the gauge-invariant intrinsic contribution that excludes the potential energy in the electric field, $\langle\tilde{\psi}_{n}(\mathbf{k})\vert\hat{V}\vert\tilde{\psi}_{n}(\mathbf{k})\rangle=e\mathbf{E}\cdot\tilde{\boldsymbol{\mathcal{A}}}_{n}$, with $\tilde{\boldsymbol{\mathcal{A}}}_{n}(\mathbf{k})
=i\langle\tilde{u}_{n}(\mathbf{k})|\partial_{\mathbf{k}}|\tilde{u}_{n}(\mathbf{k})\rangle$
the perturbed Berry connection. In contrast to the intrinsic part, Eq.~(\ref{en}) is gauge-variant due to the coordinate dependence of the potential energy.

\section{Field-perturbed velocity expectation value}
We now derive the velocity expectation value of the perturbed Bloch eigenstates. Using Eq.~(\ref{rs}), the velocity matrix elements in the unperturbed Bloch eigenstates read
\begin{equation}
 \langle\psi_{n^{\prime}}(\mathbf{k})\vert\hat{\mathbf{v}}\vert\psi_{n}(\mathbf{k})\rangle=\frac{1}{i\hbar}\langle\psi_{n^{\prime}}(\mathbf{k})\vert[\hat{\mathbf{r}},\hat{H}]\vert\psi_{n}(\mathbf{k})\rangle = \frac{1}{\hbar}\frac{\partial\varepsilon_{n}}{\partial\mathbf{k}}\delta_{nn^{\prime}}+\frac{i}{\hbar}(\varepsilon_{n^{\prime}}-\varepsilon_{n})\boldsymbol{\mathcal{A}}_{n^{\prime}n} \label{velm}
\end{equation}
which yields the velocity expectation value in the $n$th perturbed eigenstate as
\begin{equation}
\tilde{\mathbf{v}}_{n}=\langle \tilde{\psi}_{n^{\prime}}(\mathbf{k})\vert\hat{\mathbf{v}}\vert\tilde{\psi}_{n^{\prime}}(\mathbf{k})\rangle=\tilde{\mathbf{v}}_{n}^{(0)}+\tilde{\mathbf{v}}^{(1)}+\tilde{\mathbf{v}}^{(2)}+\mathcal{O}(E^3) \label{vel}
\end{equation}
where $\tilde{\mathbf{v}}_{n}^{(0)}= \langle\psi_{n}\vert\hat{\mathbf{v}}\vert\psi_{n}\rangle$, $\tilde{\mathbf{v}}_{n}^{(1)}= \langle\tilde{\psi}_{n}^{(1)}\vert\hat{\mathbf{v}}\vert\tilde{\psi}_{n}^{(0)}\rangle+\langle\tilde{\psi}_{n}^{(0)}\vert\hat{\mathbf{v}}\vert\tilde{\psi}_{n}^{(1)}\rangle$, and $ \tilde{\mathbf{v}}_{n}^{(2)}=  \langle\tilde{\psi}_{n}^{(1)}\vert\hat{\mathbf{v}}\vert\tilde{\psi}_{n}^{(1)}\rangle+\langle\tilde{\psi}_{n}^{(2)}\vert\hat{\mathbf{v}}\vert\tilde{\psi}_{n}^{(0)}\rangle+\langle\tilde{\psi}_{n}^{(0)}\vert\hat{\mathbf{v}}\vert\tilde{\psi}_{n}^{(2)}\rangle$ are the zeroth-, first-, and second-order contributions in the electric field, respectively, with $|\tilde{\psi}_{n}^{(i)}\rangle$ denoting the $i$th-order correction to the wavefunction. Substituting Eq.~(\ref{psi2}) into Eq.~(\ref{vel}), and using Eq.~(\ref{velm}), yields
\begin{align}
\tilde{v}_{n}^{a}=&\frac{1}{\hbar}\frac{\partial\varepsilon_{n}}{\partial k_{a}}+\frac{e}{\hbar}\Omega_{n}^{ab} E_{b}+\frac{e^2}{\hbar} E_{b}E_{c}\Xi_{n}^{abc}+\mathcal{O}(E^3)
\end{align}
where $\Omega_{n}^{ab}=i\sum_{m\neq n}(\mathcal{A}_{nm}^{a}\mathcal{A}_{mn}^{b}-\mathcal{A}_{nm}^{b}\mathcal{A}_{mn}^{a})=\partial_{k_{a}}\mathcal{A}_{n}^{b}-\partial_{k_{b}}\mathcal{A}_{n}^{a}$ is the Berry curvature, with its vector form given by 
$(\boldsymbol{\Omega}_{n})^{a}=\frac{1}{2}\epsilon_{abc}\Omega_{n}^{bc}$, with $\epsilon_{abc}$ the Levi-Civita symbol. The tensor $\Xi_{n}^{abc}$ describes the field-quadratic correction to the velocity. After straightforward algebra, it reduces to the form given in the main text
\begin{align}
\Xi_n^{abc}
&=\sum_{m\neq n}
\frac{\mathcal A^b_{nm}\mathcal A^c_{mn}}{(\varepsilon_m-\varepsilon_n)^2}
\partial_{k_a}(\varepsilon_m-\varepsilon_n)+i\sum_{m\neq n}
\sum_{l\neq n}\left\{\left(\frac{\mathcal A^b_{ml}\mathcal A^c_{ln}}{\varepsilon_n-\varepsilon_l}-\frac{A_{mn}^{b}A_{nn}^{c}}{\varepsilon_{n}-\varepsilon_{m}}\right)\mathcal A^a_{nm} +C.c\right\}
\nonumber\\
&\quad
+i\sum_{m\neq n}\sum_{l\neq n}
\frac{\mathcal A^b_{nm}\mathcal A^c_{ln}}{(\varepsilon_n-\varepsilon_m)(\varepsilon_n-\varepsilon_l)}
(\varepsilon_m-\varepsilon_l)\mathcal A_{ml}^a -\sum_{m\neq n}\left\{ A_{nm}^{a}\frac{\partial}{\partial k_{b}}\frac{A_{mn}^{c}}{\varepsilon_{n}-\varepsilon_{m}}+C.c\right\}
\nonumber\\
&\quad
=\sum_{m\neq n}
\frac{\mathcal A^b_{nl}\mathcal A^c_{ln}}{(\varepsilon_m-\varepsilon_n)^2}\partial_{k_a}(\varepsilon_m-\varepsilon_n)
-\sum_{l\neq n}\frac{A_{ln}^{c}\partial_{b}A_{nl}^{a}+A_{nl}^{b}\partial_{c}A_{ln}^{a}-A_{ln}^{c}\partial_{a}A_{nl}^{b}-A_{nl}^{b}\partial_{a}A_{ln}^{c}}{\varepsilon_{n}-\varepsilon_{l}}\nonumber\\
&\quad-\sum_{m\neq n}\left\{ \mathcal{A}_{nm}^{a}\frac{\partial}{\partial k_{b}}\frac{A_{mn}^{c}}{\varepsilon_{n}-\varepsilon_{m}}+C.c\right\}\nonumber\\
&\quad=\sum_{m\neq n}\left\{\frac{\partial}{\partial k_{a}}G_{nm}^{bc}-\left(\frac{\partial}{\partial k_{b}}G_{nm}^{ac} +\frac{\partial}{\partial k_{c}}G_{nm}^{ab}\right)\right\}
\end{align}  
where $G_{nm}^{ab}=\text{Re}\sum_{m\neq n}\mathcal{A}_{nm}^{a}\mathcal{A}_{mn}^{b}/(\varepsilon_{n}-\varepsilon_{m})$ is the band-normalized quantum metric.